# High-Entropy Nitride Photocatalysts for Visible-Light Antibiotic Degradation: Structural Stability, *In Situ* Interfacial Visualization, and Molecular-Level Mechanistic Insights

Zahoor Manzoor[1], Shamik Chowdhury[2]*, Raphael Benjamim de Oliveira[3], Marcelo Lopes Pereira Junior[3], Guilherme da Silva Lopes Fabris[4], Prikshat Dadhwal[5], Douglas S. Galvão[4]*, Chandra Sekhar Tiwary[6]*

[1]School of Environmental Science and Engineering,
Indian Institute of Technology Kharagpur, West Bengal 721302, India

[2]School of Water Resources,
Indian Institute of Technology Kharagpur, West Bengal 721302, India

[3]Department of Physics, Federal University of Rio Grande do Norte, Natal, RN, Brazil

[4]Applied Physics Department and Center for Computational Engineering & Sciences,
State University of Campinas, Campinas, Sao Paulo 13083-970, Brazil

[5]Hummingbird Scientific Pvt. Ltd., Hyderabad, Telangana 500016, India

[6]Department of Metallurgical and Materials Engineering,
Indian Institute of Technology Kharagpur, West Bengal 721302, India

---

† Electronic supplementary information (ESI) available.

* Corresponding author.

*E-mail*: shamikc@iitkgp.ac.in (S. Chowdhury)
chandra.tiwary@metal.iitkgp.ac.in (C. S. Tiwary)
galvao@ifi.unicamp.br (D. S. Galvão)

## ABSTRACT

High-entropy nitrides (HENs) have emerged as a promising class of advanced materials with tunable electronic structures, high stability, and abundant active sites. The incorporation of nitrogen enhances visible-light absorption, promotes efficient charge separation, and improves structural robustness, making these materials highly suitable for photocatalytic applications. In this study, (MnFeCoNiCu)N-based HEN nanoparticles (NPs) were synthesized as visible-light-assisted photocatalysts for the degradation of antibiotics, including sulfamethoxazole (SME) and tetracycline (TCL). The catalyst exhibited excellent performance, achieving 96% degradation of SME and 94% removal of TCL within 2 h of visible-light irradiation. The photocatalytic activity was systematically evaluated under varying operational parameters, including solution pH, catalyst dosage, pollutant concentration, and the presence of coexisting ions. Notably, the catalyst maintained high efficiency in real water matrices, demonstrating its practical applicability. The entropy-stabilized nitride framework exhibited negligible metal leaching, excellent thermal and structural stability as confirmed by temperature-dependent synchrotron angle-dispersive X-ray diffraction, and stable performance over multiple reuse cycles. Furthermore, in situ liquid-cell transmission electron microscopy provided real-time insight into catalyst–pollutant interactions, while complementary molecular simulations revealed the structural stability of the HEN NPs during molecular adsorption and the distinct interaction modes of the two antibiotics. Phytotoxicity tests using *Vigna radiata* confirmed the effective detoxification of treated solutions. Overall, this work establishes (MnFeCoNiCu)N HENs as efficient and durable visible-light-driven photocatalysts for the removal of antibiotics from complex aqueous environments.

## 1. Introduction

The widespread occurrence of antibiotics in aquatic environments has become a global environmental concern due to their extensive consumption, incomplete metabolism, and continuous discharge from domestic, hospital, agricultural, and pharmaceutical sources [1]. Their persistence in water bodies promotes the proliferation of antibiotic-resistant microorganisms and poses significant risks to aquatic ecosystems and human health [2,3]. Conventional wastewater treatment processes are generally ineffective in completely removing these recalcitrant contaminants [4], highlighting the need for efficient and sustainable remediation technologies. Among the various advanced treatment strategies, semiconductor-based photocatalysis has attracted considerable attention because it enables the degradation of persistent organic contaminants under mild operating conditions without the continuous addition of chemical oxidants [5,6]. The development of advanced photocatalytic materials with high activity, excellent stability, and broad solar-light utilization therefore remains an active area of research.

Recent years have witnessed tremendous interest in high-entropy materials (HEMs), a class of multicomponent materials consisting of multiple principal elements arranged within a single crystalline lattice. Unlike conventional materials, HEMs derive their unique physicochemical properties from high configurational entropy, severe lattice distortion, sluggish diffusion, and synergistic interactions among constituent elements [7]. These characteristics impart exceptional structural stability, tunable electronic structures, abundant catalytically active sites, and remarkable chemical durability, making HEMs promising candidates for numerous catalytic, electrochemical, and environmental applications [8,9]. Consequently, diverse HEM families, including high-entropy alloys, oxides, sulfides, carbides, borides, and nitrides, have been explored for applications ranging from energy conversion and storage to environmental remediation [10,11].

Despite these advances, investigations have predominantly focused on high-entropy alloys and oxides, whereas several other HEM families remain comparatively underexplored. Extending research to high-entropy carbides, nitrides, borides, sulfides, and related compounds is essential for developing a comprehensive understanding of the structure–property relationships governing this emerging class of materials. Each HEM family possesses distinct local bonding characteristics, electronic structures, and lattice disorder, resulting in unique physicochemical properties and application potential [12]. For example, high-entropy carbides exhibit exceptional resistance to oxidation and wear [13], whereas high-entropy nitrides (HENs) combine outstanding chemical stability with excellent solar-energy harvesting capability [14,15]. These characteristics make HENs particularly attractive for environmental applications, where catalyst durability, broad spectral response, and long-term operational stability are of paramount importance. Nevertheless, the potential of HENs for the removal of emerging contaminants, particularly antibiotics, remains largely unexplored.

Additionally, understanding the dynamic interactions occurring at the catalyst–pollutant interface is essential for the rational design and optimization of next-generation HEMs. However, mechanistic investigations of HEN-mediated degradation processes remain scarce. Recent advances in *in situ* liquid-cell transmission electron microscopy (LCTEM) enable the direct visualization of catalyst–pollutant interactions under realistic liquid-phase conditions, providing unprecedented insights into interfacial processes and structural evolution [16,17]. When integrated with molecular simulations, these techniques provide a powerful framework for correlating atomic-scale interactions with experimentally observed degradation pathways, catalyst stability, and overall catalytic performance [18].

Based on these considerations, (MnFeCoNiCu)N nanoparticles (NPs) were synthesized through induction melting followed by ball milling and evaluated as visible-light-driven photocatalysts for the degradation of the representative antibiotics sulfamethoxazole (SME) and tetracycline (TCL). The effects of key operational parameters, including solution pH, catalyst dosage, pollutant concentration, and coexisting ions, were systematically investigated, while catalyst stability was assessed through repeated cycling experiments, metal-leaching analysis, and temperature-dependent synchrotron angle-dispersive X-ray diffraction (ADXRD). The practical applicability of the catalyst was further evaluated in real water matrices, and the detoxification of the treated solutions was verified through phytotoxicity assays using *Vigna radiata*. Furthermore, *in situ* LCTEM, together with complementary molecular simulations, was employed to elucidate the interactions between HEN nanoparticles and antibiotic molecules, thereby providing molecular-level insights into the degradation mechanism. To the best of our knowledge, this study represents the first comprehensive investigation of HENs for visible-light-driven antibiotic degradation, integrating photocatalytic performance, structural stability, real-time interfacial visualization, molecular simulations, and ecotoxicity assessment within a unified framework.

## 2. Materials and methods

### 2.1. Materials

Details of the chemicals, reagents, and other materials used in this study are provided in the Supplementary Information.

### 2.2. Synthesis of (MnFeCoNiCu)N NPs

MnFeCoNiCu high-entropy alloy (HEA) nanoparticles (NPs) were synthesized via induction melting, followed by annealing and ball milling, using high-purity Mn, Fe, Co, Ni, and Cu metal chips as precursors. An equimolar mixture of the metals was melted at 1250 °C for 10 minutes in an induction furnace under an inert atmosphere. The resulting alloy ingot was subsequently homogenized by annealing at 1000 °C for 96 h in a muffle furnace. The homogenized alloy was then subjected to planetary ball milling for 12 h to produce MnFeCoNiCu HEA NPs. The synthesized HEA NPs were subsequently mixed with melamine at HEA: melamine molar ratios of 1:2.5, 1:5, and 1:10 to prepare precursors for the corresponding HEN NPs. Nitridation was carried out by heat treatment at 800 °C in a quartz tube under an argon atmosphere, using controlled heating and cooling rates of 5 °C $min^{-1}$. The resulting samples were designated as HEN-1:2.5, HEN-1:5, and HEN-1:10 according to their respective precursor ratios.

### 2.3. Characterization of (MnFeCoNiCu)N NPs

Detailed descriptions of the characterization techniques employed for the synthesized (MnFeCoNiCu)N NPs are provided in the Supplementary Information.

### 2.4. Photocatalytic experiments

The photocatalytic performance of the synthesized (MnFeCoNiCu)N NPs was evaluated for the degradation of SME and TCL in aqueous solution at 25 °C. All experiments were performed in a 250 mL triple-jacketed quartz immersion-type batch photoreactor (Lelesil Innovative Systems, India) under visible-light irradiation. The reactor was equipped with a 250 W high-pressure mercury vapor lamp fitted with a UV cut-off filter to eliminate wavelengths below 420 nm. Further details of the experimental procedures are provided in the Supplementary Information.

### 2.5. *In situ* LCTEM study

*In situ* LCTEM was performed using a 1400 Series liquid electrochemistry TEM holder (Hummingbird Scientific, USA) coupled with a JEM-ARM200F NEOARM atomic-resolution analytical transmission electron microscope (JEOL Ltd., Japan). For sample preparation, HEN-1:5 NPs were dispersed in isopropanol (IPA), and approximately 2 μL of the suspension was drop-cast onto a silicon nitride chip (Hummingbird Scientific) containing a 50 nm membrane and a 200 μm × 50 μm observation window. The chip was dried under ambient conditions before assembly of the liquid cell. Following assembly, the liquid cell was examined for leakage using a dry vacuum pump while deionized (DI) water was circulated through the system. After confirming proper sealing, the TEM holder was inserted into the microscope. During in situ imaging, 5 mg $L^{-1}$ TCL solution was introduced into the liquid cell using a syringe pump at a controlled flow rate of 2 μL $min^{-1}$.

## 2.6. ADXRD study

Temperature-dependent synchrotron powder diffraction were performed using the BL-12 ADXRD beamline at the 2.5 GeV Indus-2 synchrotron radiation facility, Raja Ramanna Centre for Advanced Technology, Indore, India. The incident photon energy was calibrated using a National Institute of Standards and Technology (NIST) standard lanthanum hexaboride ($LaB_6$) reference material, resulting in an X-ray wavelength of 0.72 Å. Diffraction patterns were collected over a 2θ range of 10°–55° to monitor temperature-induced structural evolution. The high brilliance, monochromaticity, and angular resolution of the synchrotron radiation enabled accurate analysis of crystal structure, phase stability, and lattice distortions under varying thermal conditions.

## 2.7. Theoretical Methodology

First-principles calculations were performed within the framework of Density Functional Theory (DFT), using the SIESTA code [19,20]. Exchange–correlation effects were described using the Perdew–Burke–Ernzerhof (PBE) functional within the generalized gradient approximation (GGA) [21]. For Fe-containing systems, the on-site Coulomb interaction associated with the partially localized Fe 3d electrons was treated using the DFT+U formalism. A Hubbard parameter of U = 4.0 eV was applied to the Fe 3d orbitals, consistent with values commonly reported for Fe-based compounds [22].

Electronic wave functions were expanded using a double-ζ polarized numerical atomic-orbital basis set, while the electronic density was represented on a real-space grid with a mesh cutoff energy of 350 Ry. Brillouin zone sampling employed Γ-centered Monkhorst-Pack **k**-point meshes [23], with mesh densities optimized for each crystalline system to ensure convergence of the total energy and structural properties. Structural optimizations were performed until the maximum residual force on each atom was below 0.05 eV $Å^{-1}$, and electronic self-consistency was achieved when the difference between successive density matrices was less than $10^{-4}$.

To investigate the influence of N incorporation on the structural and electronic properties of the, a dataset comprising 100 statistically independent equimolar MnFeCoNiCu configurations was generated. The reference face-centered cubic (FCC) lattice containing 72 metallic sites was generated using the Atomic Simulation Environment (ASE) package [24]. Chemical disorder was introduced by randomly distributing the constituent elements over the metallic sublattice while maintaining an equimolar composition, thereby generating statistically independent atomic configurations representative of chemically disordered HEAs. The number of atoms considered for each composition is summarized in Table S1.

A vacuum layer of 30 Å was introduced along the *z*-direction to eliminate interactions between periodic images. Structural relaxations involved the simultaneous optimization of atomic coordinates and lattice parameters until the maximum residual forces was below 0.05 eV $Å^{-1}$, while identical electronic convergence criteria were maintained for all systems. To obtain a representative model of the experimentally synthesized material, the lowest-energy HEA configuration was selected as the precursor for nitrogen incorporation. N atoms were subsequently introduced into the optimized HEA structure, yielding a chemically ordered 105-atom supercell containing 32 N atoms.

Optical properties were investigated through additional ab initio calculations, using the same computational parameters and convergence criteria described above. Brillouin-zone sampling was refined using an 8×8×1 Monkhorst–Pack **k**-point mesh, while the 30 Å vacuum layer was retained along the *z*-direction to avoid spurious interactions between periodic images. These computational settings ensured converged electronic structures and reliable optical properties.

The optical absorption coefficient (α) was calculated as a function of the photon energy (ω) according to [25].

$$\alpha(\omega) = \sqrt{2}\,\omega\left[\left(\varepsilon_1^2(\omega) + \varepsilon_2^2(\omega)\right)^{1/2} - \varepsilon_1(\omega)\right]^{1/2} \qquad (1)$$

where $\varepsilon_1(\omega)$ and $\varepsilon_2(\omega)$ are the real and imaginary components of the dielectric function, respectively.

The adsorption behavior of SME and TCL on the HEN surface was investigated through atomistic simulations using the Density Functional Tight-Binding (DFTB) approach combined with the extended tight-binding (xTB) method [26] under the GFN1-xTB parametrization [27], as implemented in the DFTB+ package. To capture the compositional and structural complexity of the HEN surface, a 2×2×1 supercell was employed. Structural optimizations were performed using convergence thresholds of $10^{-5}$ Ha for both the total energy and self-consistent charge (SCC) iterations, while the maximum absolute gradient was limited to $10^{-4}$ Ha Bohr$^{-1}$. Brillouin-zone integration was carried out using Γ-point sampling.

A Monte Carlo sampling approach was first employed to identify energetically favorable adsorption sites on the chemically disordered HEN surface. A large number of adsorption configurations were generated and ranked according to their total energies, enabling the identification of the most energetically favorable adsorption zones. These regions consistently corresponded to local atomic environments with higher nitrogen coordination and greater exposure of metallic active sites.

The lowest-energy adsorption configurations were subsequently fully optimized to obtain the equilibrium geometries. The molecule–surface separation was further varied by ±0.2 Å along the surface normal (*z*-direction) to determine the equilibrium adsorption distance. The binding energy ($E_b$) was then calculated as

$$E_b = E_{\mathrm{HEN+Molecule}} - \left( E_{\mathrm{HEN}} + E_{\mathrm{Molecule}} \right) \quad (2)$$

where $E_{\mathrm{HEN+Molecule}}$ is the total energy of the HEN surface with the adsorbed molecule (SME or TCL), $E_{\mathrm{HEN}}$ is the energy of the isolated substrate, and $E_{\mathrm{Molecule}}$ is the energy of the isolated adsorbate.

## 3. Results and discussion

### 3.1. Characterization of (MnFeCoNiCu)N NPs

The morphology and microstructure of the as-synthesized (MnFeCoNiCu)N NPs were comprehensively characterized using atomic force microscopy (AFM), electron microscopy, and X-ray diffraction. AFM images (Fig. 1a,b) revealed that the NPs possessed an average thickness of approximately 56.8 nm and a lateral dimension of about 0.20 μm. Field-emission scanning electron microscopy (FESEM) analysis (Fig. 1c) further confirmed the formation of NPs with a relatively uniform surface morphology and surface texture. Bright-field transmission electron microscopy (TEM) imaging (Fig. 1d) revealed well-defined NPs without discernible structural defects or distortions, while the uniform image contrast suggests a homogenous distribution of the constituent elements throughout the particles. The crystalline nature of the (MnFeCoNiCu)N NPs was confirmed by the high-resolution TEM (HRTEM) image (Fig. 1e) together with its corresponding fast Fourier transform (FFT) pattern (Fig. 1f). Well-resolved lattice fringes with an interplanar spacing of approximately 0.21 nm were observed (Fig. 1g), corresponding to the (111) plane of a face-centered (FCC) crystal structure. Furthermore, the selected area electron diffraction (SAED) pattern (Fig. 1h) exhibited well-defined diffraction rings without any additional spots, confirming the formation of a single-phase solid solution. Scanning TEM (STEM) energy-dispersive X-ray spectroscopy (EDS) elemental mapping (Fig. 1i–o) further verified the successful synthesis of the HEN by demonstrating the homogenous distribution of Mn, Fe, Co, Ni, Cu, and N throughout the NPs, with no evidence of elemental segregation.

The crystal structure of the synthesized HEN NPs was further examined by X-ray diffraction (XRD), and the diffraction pattern is presented in Fig. 2a. A weak diffraction peak centered at approximately 26.1° was assigned to the (002) plane of graphitic carbon [28], originating from residual carbon species generated during the nitridation process. The diffraction peaks, located at 2θ values of approximately 35.5°, 41.2°, 43.3°, 59.7°, 71.3°, and 74.9°, can be indexed to the (111), (200), (200), (220), (311), and (222) crystallographic planes of FCC structure [29,30], respectively. In addition, the diffraction peak observed at approximately 50.8° agrees well with the standard cubic transition-metal nitride phase (JCPDS No. 01-075-2127; space group $Fm\overline{3}m$ [31]. The absence of secondary crystalline phases, together with the well-defined diffraction peaks, confirms the successful formation of a crystalline single-phase HEN. The stabilization of the FCC structure is attributed to the strong affinity of the constituent transition metals toward N, which promotes the formation of a stable multicomponent nitride solid solution.

Subsequently, Fourier transform infrared (FTIR) spectroscopy was employed to investigate the surface chemistry of the (MnFeCoNiCu)N NPs (Fig. 2b). The FTIR spectrum exhibits characteristic vibrational bands associated with metal–oxygen bonds, indicating partial surface oxidation of the constituent transition metals upon exposure to ambient conditions. Specifically, the bands at 503 and 618 $cm^{-1}$ are assigned to Cu–O stretching vibrations [32,33], while those at 520 and 538 $cm^{-1}$ correspond to Mn–O stretching modes [34,35]. The absorption bands centered at 530, 562, and 607 $cm^{-1}$ are attributed to Fe–O [36], Co–O [37], and Ni–O [38] stretching vibrations, respectively. The presence of these oxide-related bands suggests the formation of a thin native oxide layer on the NP surface, which is commonly observed for transition-metal-based materials exposed to air.

The surface chemical composition and oxidation states of the constituent elements were further examined by X-ray photoelectron spectroscopy (XPS). The survey spectrum (Fig. S1) confirms the presence of Mn, Fe, Co, Ni, Cu, N, and O. The distinct O 1s signal corroborates the formation of a surface metal oxide layer inferred from the FTIR analysis. High-resolution XPS spectra (Fig. 2g–l) reveal the coexistence of metallic and oxidized species for the constituent transition metals. The Mn 2p XPS spectrum (Fig. 2g) exhibit peaks at 641.6 and 653.7 eV, corresponding to $Mn^{3+}$ $2p_{3/2}$ and $Mn^{3+}$ $2p_{1/2}$, respectively [39,40], along with a satellite peak at 644.5 eV associated with Mn−O bonding [40]. The Fe 2p spectrum (Fig. 2h) display peaks at 712.2 and 715.2 eV corresponding to $Fe^{3+}$ ($2p_{3/2}$) species [41,42], while the peaks at at 721.2 and 727.5 eV are assigned to $Fe^{3+}$ $2p_{1/2}$ [42,43]. A weaker peak located at 720.5 eV is attributed to metallic Fe [44], indicating that both metallic and oxidized Fe species coexist on the NP surface. Similarly, the Co 2p spectrum (Fig. 2i) presents characteristic peaks at 778.4 and 780.5 eV, corresponding to $Co^{3+}$ and $Co^{2+}$ $2p_{3/2}$ species, respectively [45]. In the Ni 2p spectrum (Fig. 2j), the peaks centered at 854.2 and 872.5 eV are assigned to $Ni^{2+}$ $2p_{3/2}$ and $Ni^{2+}$ $2p_{1/2}$, respectively [41], accompanied by two characteristic satellite peaks at 861.2 and 880.1 eV [43]. The Cu 2p spectrum (Fig. 2k) comprise peaks at 932.6 and 952.4 eV, corresponding to $Cu^{+}$ $2p_{3/2}$ and $Cu^{+}$ $2p_{1/2}$ [46], together with peaks at 934.5 and 953.8 eV assigned to $Cu^{2+}$ $2p_{3/2}$ and $Cu^{2+}$ $2p_{1/2}$ [46,47]. A weak shake-up satellite centered at approximately 962.2 eV further confirms the presence of $Cu^{2+}$ species [48]. The deconvoluted N 1s specturm (Fig. 2l) can be resolved into three components centered at 398.6, 401.1, and 403.2 eV, corresponding to pyridinic N, quaternary N, and oxidized N–O species, respectively [49,50]. Overall, the XPS results confirm the successful incorporation of N into the HEN framework, while also revealing the coexistence of multiple oxidation states of the constituent transition metals, consistent with the formation of a chemically complex HEN surface. The presence of mixed-valence metal species and surface N-functionalities is expected to facilitate charge transfer and provide abundant catalytically active sites, thereby contributing to the enhanced photocatalytic performance of the synthesized HEN NPs.

Additionally, the optical and photoelectrochemical properties of the synthesized HEN NPs were systematically investigated to elucidate their visible-light response and charge-transfer characteristics. UV–visible diffuse reflectance spectroscopy recorded strong and broad absorption throughout the visible region for all HEN samples (Fig. 2c). Among them, HEN-1:5 exhibited the highest absorption intensity, indicating superior visible-light harvesting capability compared with HEN-1:2.5 and HEN-1:10. The enhanced optical response is expected to facilitate the generation of a greater number of photogenerated charge carriers under visible-light irradiation. The optical bandgap ($E_g$) of HEN-1:5 was estimated from the Kubelka–Munk plot to be 1.72 eV (Fig. 2d), confirming its narrow-band-gap semiconducting nature and excellent suitability for visible-light-driven photocatalysis. Such a narrow band gap enables efficient utilization of the solar spectrum while promoting the generation of photoexcited electrons and holes. Electrochemical impedance spectroscopy (EIS) was employed to evaluate the charge transfer behavior of the synthesized HENs. As shown in the Nyquist plots (Fig. 2e), HEN-1:5 exhibits the smallest semicircle diameter among the three samples, indicating the lowest charge-transfer resistance and the most efficient separation and migration of photogenerated charge carriers. The reduced interfacial resistance suppresses electron–hole recombination, thereby enhancing the photocatalytic activity of HEN-1:5. The semiconductor characteristic of HEN-1:5 was further examined using Mott–Schottky analysis (Fig. 2f). The positive slope of the Mott–Schottky plot confirms the n-type semiconducting behavior of the material. The flat-band potential ($E_{fb}$) was determined to be 1.53 V (vs. the standard hydrogen electrode (SHE)). As the conduction band potential ($E_{CB}$) for n-type semiconductors is typically 0.1–0.2 V more negative than $E_{fb}$ [51,52], the $E_{CB}$ of HEN-1:5 was estimated to be approximately 1.43 V (vs. SHE). Using the relationship $E_{VB} = E_{CB} + E$g, the valence band potential ($E_{VB}$) was calculated to be 3.15 V (vs. SHE). The favorable band-edge alignment, together with the narrow band gap and efficient charge separation, provides the thermodynamic driving force for the generation of reactive oxygen species (ROS) under visible-light irradiation.

### 3.2. Photocatalytic performance of (MnFeCoNiCu)N NPs

Fig. 3a,b compares the visible-light photocatalytic performance of the synthesized HENs toward the degradation of SME and TCL. Negligible adsorption of both antibiotics was observed during the initial 30 min dark-equilibrium period, confirming that the subsequent pollutant removal under visible-light irradiation was predominantly governed by photocatalytic oxidation rather than adsorption. The photocatalytic experiments were performed at an initial solution pH of approximately 7.0, which remained essentially unchanged after the addition of the HENs. In the absence of the photocatalyst, only 36% of SME and 31% of TCL were degraded after 120 min of visible-light irradiation, demonstrating that direct photolysis alone is insufficient for efficient antibiotic degradation. In contrast, all synthesized HENs exhibited substantially enhanced photocatalytic activity, with HEN-1:5 demonstrating the best performance by achieving degradation efficiencies of 96% for SME and 94% for TCL within 120 min. The photocatalytic activity followed the order HEN-1:5 > HEN-1:10 > HEN-1:2.5, corresponding to degradation efficiencies of 96%, 73%, and 55% for SME, and 94%, 62%, and 53% for TCL, respectively. The superior photocatalytic performance of HEN-1:5 is attributed to its enhanced visible-light absorption, narrow band gap, and reduced charge-transfer resistance, which collectively promote efficient light harvesting, suppress charge-carrier recombination, and accelerate interfacial electron transfer. Furthermore, the incorporation of N into the high-entropy lattice promotes strong metal–N bonding and hybridization between transition-metal *d* orbitals and N 2*p* orbitals, thereby further enhancing visible-light absorption and charge separation [53,54].

To identify the reactive species responsible for the photocatalytic degradation of SME and TCL, selective scavenging experiments were performed using isopropyl alcohol (IPA), ascorbic acid (AA), potassium iodide (PI), and sodium azide (SA) as scavengers for hydroxyl radical (•OH), superoxide anion ($•O_2^-$), VB holes ($h^+_{VB}$), and singlet oxygen ($^1O_2$), respectively. As shown in Fig. 3c,d, the addition of IPA resulted in the most pronounced suppression of photocatalytic activity, reducing the degradation efficiencies to 59% for SME and 55% for TCL. These results indicate that •OH radicals are the dominant reactive species governing antibiotic degradation. A substantial decrease in degradation efficiency was also observed in the presence of PI, demonstrating that photogenerated $h^+_{VB}$ actively participate in the oxidation process. In comparison, the addition of SA produced only a marginal reduction in photocatalytic activity, suggesting that $^1O_2$ plays a negligible role under the present reaction conditions.

Based on the radical scavenging experiments and the experimentally determined band-edge positions, a plausible visible-light-driven photocatalytic degradation mechanism is proposed (Fig. 4). Upon visible-light irradiation, HEN-1:5 generates electron–hole pairs. The highly positive $E_{VB}$ enables photogenerated holes to oxidize surface hydroxide ions and water molecules ($H_2O$), producing •OH radicals. These highly reactive species subsequently attack SME and TCL molecules, resulting in their progressive oxidation and eventual mineralization into carbon dioxide ($CO_2$), $H_2O$, and other low-molecular-weight products. The proposed mechanism is in good agreement with both the ROS trapping experiments and the experimentally determined electronic band structure.

The excellent photocatalytic performance of HEN-1:5 was further validated through kinetic and mineralization analyses. The degradation kinetics followed the Langmuir–Hinshelwood model (Fig. S2a,b), with HEN-1:5 exhibiting the highest apparent first-order rate constants for both SME and TCL among the synthesized HENs, confirming its rapid degradation capability. Residual total organic carbon analysis (Fig. S3) further demonstrated mineralization efficiencies of 83% for SME and 78% for TCL, indicating that the parent antibiotics were extensively mineralized rather than merely transformed into intermediate products. To further assess the effectiveness of the synthesized HEN, its photocatalytic performance was compared with those of recently reported visible-light-driven photocatalysts for antibiotic degradation (Table S2). HEN-1:5 exhibits competitive degradation efficiency under comparatively mild operating conditions, highlighting the potential of high-entropy nitrides as a promising class of visible-light photocatalysts for the efficient removal of emerging pharmaceutical contaminants.

To gain molecular-level insight into the photocatalytic degradation process, the transformation pathways of SME and TCL were elucidated from the intermediates identified by liquid chromatography–mass spectrometry (LC–MS) analysis. The detected intermediates indicate that both antibiotics undergo successive oxidative transformations involving hydroxylation, oxidation, bond cleavage, demethylation, and ring-opening reactions before their eventual mineralization into inorganic end products. For SME (*m*/*z* = 254), three major degradation pathways were identified (Fig. 5). In Pathway I, •OH radical attack on the isoxazole ring initially produces the hydroxylated intermediate S1 (*m*/*z* = 275), which subsequently undergoes S–N bond cleavage to generate S2 (*m*/*z* = 135) [55]. In Pathway II, ROS-induced cleavage of the sulfonamide linkage yields S3 (*m*/*z* = 156) and S4 (*m*/*z* = 99) [56], reflecting the sulfonamide sulfur atom's susceptibility to oxidative attack [57]. Subsequent hydroxylation of S3 forms S5 (*m*/*z* = 108). In Pathway III, SME undergoes denitration and hydroxylation to generate an intermediate at *m*/*z* = 178, followed by desulfonation to produce S4 (*m*/*z* = 99) [58]. The progressive cleavage of the isoxazole and sulfonamide moieties, together with successive oxidation reactions, ultimately converts SME into low-molecular-weight intermediates, which are further mineralized into $CO_2$, $H_2O$, and inorganic ions, including ammonium ($NH_4^+$), nitrate ($NO_3^-$), and sulfate ($SO_4^{2-}$).

The proposed degradation pathways for TCL (*m*/*z* = 445) are illustrated in Fig. 6. Under continuous visible-light irradiation, TCL undergoes a series of oxidative transformations involving demethylation, deamination, hydroxylation, and sequential ring-opening reactions [59,60]. In Pathway I, successive demethylation and deamination reactions generate intermediates T1 (*m*/*z* = 417), T2 (*m*/*z* = 402), and T3 (*m*/*z* = 359), followed by C–C bond cleavage to form the lower-molecular-weight fragments T4 (*m*/*z* = 193) and T5 (*m*/*z* = 173) [61]. Pathway II proceeds through substitution and dihydroxylation reactions, producing intermediates T6 (*m*/*z* = 467), T7 (*m*/*z* = 451), and T8 (*m*/*z* = 389) [62]. In Pathway III, *N*-demethylation initially forms T9 (*m/z* = 427), which subsequently undergoes deamination and oxidation to produce T10 (*m/z* = 415), T11 (*m/z* = 371), and T12 (*m/z* = 327). Further oxidative cleavage results in extensive fragmentation of the TCL framework, generating T13 (*m/z* = 149) together with smaller intermediates, including T14 (*m/z* = 167), T15 (*m/z* = 126), T16 (*m/z* = 123), T17 (*m/z* = 116), and T18 (*m/z* = 102) [63]. These intermediates are subsequently converted into low-molecular-weight products and finally mineralized into $CO_2$, $H_2O$, $NH_4^+$, and $NO_3^-$.

The proposed degradation pathways are consistent with the ROS scavenging experiments, which identified •OH radicals and photogenerated $h^+_{VB}$ as the dominant reactive species responsible for antibiotic degradation. The extensive oxidation, bond cleavage, and ring-opening reactions observed for both antibiotics explain the high mineralization efficiencies achieved by HEN-1:5, demonstrating its capability to effectively degrade and detoxify antibiotic-contaminated water.

Although HEN-1:5 exhibited excellent photocatalytic activity toward SME and TCL degradation, its performance under different operating conditions remains equally important for practical applications. Therefore, the effects of key operational parameters, including solution pH, catalyst dosage, initial pollutant concentration, and coexisting inorganic ions, on the photocatalytic degradation of SME and TCL were systematically investigated.

Among these parameters, solution pH plays a particularly important role because it governs both the surface charge of the photocatalyst and the aqueous speciation of the target pollutants, thereby influencing adsorption behavior, ROS generation, and the overall photocatalytic efficiency. The point of zero charge (pHpzc) of the HEN-1:5 was determined to be approximately 3.35 (Fig. S4), indicating that the catalyst surface is positively charged below pH 3.35 and negatively charged above this value. Meanwhile, SME possesses two acid dissociation constants (p$K_{a1}$ = 1.6 and p$K_{a2}$ = 5.7), existing predominantly as a cation below pH 1.6, a zwitterion between pH 1.6 and 5.7, and an anion above pH 5.7 [64]. Accordingly, relatively high degradation efficiencies of 89% and 91% were achieved at pH 3 and 5, respectively (Fig. 7a), owing to favorable electrostatic interactions between SME and the HEN-1:5 surface. Interestingly, the degradation efficiency increased further to 96% at pH 7 despite SME predominantly existing in its anionic form. This observation suggests that, under near-neutral conditions, the enhanced generation and reactivity of ROS outweigh electrostatic effects. At pH ≥ 8, however, the degradation efficiency declined because electrostatic repulsion between the negatively charged catalyst surface and anionic SME molecules became dominant

Unlike SME, TCL exhibits more complex acid–base behavior with three acid dissociation constants (p$K_{a1}$ = 3.3, p$K_{a2}$ = 7.7, and p$K_{a3}$ = 9.7) [65,66]. The highest degradation efficiency (89%) was obtained at pH 7 (Fig. 7b), where TCL predominantly exists in its zwitterionic form, favoring interaction with the HEN-1:5 surface. The degradation efficiency decreased slightly at pH 5 and 9 because of weaker electrostatic interactions, while a more pronounced reduction was observed at pH 3 (51%) and pH 11 (43%). Under strongly alkaline conditions, electrostatic repulsion, together with the consumption of photogenerated holes by excess hydroxide ions, suppresses ROS generation and consequently limits photocatalytic degradation [67].

Catalyst dosage is another important operational parameter because it governs the number of available active sites and, consequently, the overall photocatalytic efficiency. As shown in Fig. 7c,d, increasing the HEN-1:5 dosage from 0.10 to 0.50 g $L^{-1}$ significantly enhanced the degradation of both antibiotics, with SME removal increasing from 75% to 96% and TCL removal from 65% to 94%. This improvement is attributed to greater availability of catalytically active sites and increased ROS generation. However, further increasing the catalyst dosage beyond 0.50 g $L^{-1}$ resulted in a decline in degradation efficiency. At higher catalyst loadings, increased suspension turbidity and particle agglomeration reduce light penetration through enhanced photon scattering and absorption, thereby limiting charge-carrier generation and decreasing the accessibility of active sites.

The influence of the initial antibiotic concentration on photocatalytic degradation is presented in Fig. 7e,f. Increasing the pollutant concentration from 5 to 20 mg $L^{-1}$ reduced the degradation efficiency from 96% to 63% for SME and from 94% to 59% for TCL. Under constant irradiation and catalyst loading, the amount of ROS generated remains essentially unchanged and becomes insufficient to completely oxidize the larger number of pollutant molecules present at higher concentrations. Moreover, the accumulation of degradation intermediates competes with the parent antibiotics for reactive species and catalytically active sites, further suppressing the overall degradation efficiency.

The robustness of HEN-1:5 was further evaluated in the presence of common inorganic anions frequently encountered in natural and wastewater matrices, including chloride ($Cl^{-}$), $SO_4^{2-}$, hydrogen phosphate ($HPO_4^{2-}$), $NO_3^{-}$, and bicarbonate ($HCO_3^{-}$). As shown in Fig. 7g,h, all tested anions inhibited the photocatalytic degradation of both antibiotics to varying extents. For SME, the inhibitory effect followed the order $HCO_3^{-} > HPO_4^{2-} > SO_4^{2-} > Cl^{-} > NO_3^{-}$, whereas for TCL the order was $HPO_4^{2-} > HCO_3^{-} > SO_4^{2-} > NO_3^{-} > Cl^{-}$. The pronounced inhibition by $HCO_3^{-}$ and $HPO_4^{2-}$ is primarily attributed to their strong scavenging ability toward •OH radicals and photogenerated electrons, resulting in the formation of less reactive species such as $•CO_3^{-}$, $•CO_2^{-}$, and $•HPO_4^{-}$. In addition, the buffering action of these anions can modify the local solution chemistry, thereby suppressing ROS generation and reducing photocatalytic activity [68,69]. Although $SO_4^{2-}$, $NO_3^{-}$, and $Cl^{-}$ also consume •OH radicals, their inhibitory effects are comparatively less pronounced because of their lower radical-scavenging capacities.

### 3.3. *In situ* LCTEM study

Figure 8 presents a time-resolved sequence of LCTEM images as well as a schematic representation of the dynamic interactions between HEN-1:5 NPs and TCL molecules under liquid-phase conditions. Immediately after the introduction of the 5 mg $L^{-1}$ TCL solution into the liquid cell (Fig. 8i), the solution flows across the HEN-1:5 surface (Fig. 8i–ii), initiating interfacial interactions between the catalyst and pollutant molecules (Supplementary Video V1). As the interaction progresses, TCL molecules preferentially associate with the NP edges, suggesting that these regions possess a higher density of energetically favorable active sites. With increasing contact time, the surface coverage gradually extends across the entire NP, resulting in a more uniform distribution of TCL molecules (Fig. 8iii). After the TCL flow was terminated, the associated molecules remained localized at the HEN-1:5 surface, indicating persistent catalyst–pollutant interactions under liquid-phase conditions. Importantly, no discernible changes in the size, morphology, or structural integrity of the NPs were observed throughout the experiment (Fig. 8iv). The excellent structural stability, together with the sustained catalyst–pollutant interfacial interactions, demonstrates the robustness of HEN-1:5 under reaction conditions and provides direct experimental evidence of its ability to facilitate heterogeneous photocatalytic reactions.

### 3.4. ADXRD study

To investigate the structural stability of HEN-1:5 and determine whether interaction with SME induces any crystallographic changes, temperature-dependent ADXRD measurements were performed on pristine HEN-1:5 and the HEN-1:5–SME system. The corresponding ADXRD patterns are presented in Fig. S5a,b. For pristine HEN-1:5, the ADXRD patterns exhibit sharp and well-defined diffraction peaks throughout the investigated temperature range, confirming the highly crystalline nature of the material. As the temperature increases, the diffraction peaks gradually shift toward lower 2θ values, consistent with thermal expansion of the crystal lattice. Importantly, no peak splitting, disappearance of reflections, or emergence of additional diffraction peaks is observed, indicating the absence of any crystallographic phase transformation during heating. The retention of the diffraction features throughout the temperature range demonstrates that the crystal symmetry and phase purity of the HEN remain unaffected, confirming its excellent thermal stability.

The HEN-1:5–SME system exhibits a similar crystallographic response, with all characteristic reflections of the HEN phase remaining clearly preserved throughout the investigated temperature range. This observation indicates that interaction with SME does not alter the fundamental crystal structure of the HEN. Compared with pristine HEN-1:5, however, HEN-1:5–SME displays slight peak broadening along with minor variations in diffraction intensity as the temperature increases, suggesting localized lattice strain, interfacial interactions, or defect generation associated with the adsorbed SME molecules. The absence of additional diffraction peaks further indicates that SME remains either amorphous or highly dispersed on the HEN surface without forming a separate crystalline phase. Moreover, the slightly greater peak shift suggests a modest change in thermal expansion behaviour arising from interactions between the catalyst and the adsorbed antibiotic.

These results demonstrate that the HEN framework retains its crystallographic integrity both before and after interaction with SME, without undergoing any detectable phase transformation. The excellent thermal stability and resistance to structural modification upon pollutant adsorption highlight the robustness of HEN-1:5, supporting its suitability for repeated and long-term photocatalytic wastewater treatment applications.

### 3.5. Theoretical Analysis

The analysis commenced with the structural optimization of all 100 HEA configurations. To identify suitable precursors for N incorporation, the optimized structures were ranked according to their total energies. During structural relaxation, the lattice parameters were allowed to vary freely. In general, the MnFeCoNiCu HEA structures exhibited pronounced distortions, with at least one lattice angle varying by 10-15%. This behavior is consistent with the atomic size and electronegativity mismatch among the constituent elements, which induces local charge redistribution, modifies atomic packing, and promotes structural relaxation [70,71]. In particular, the presence of Cu is expected to further enhance these effects by favoring local compositional heterogeneity and lattice strain. The present analysis primarily focuses on the lattice angles because they are more sensitive to local structural distortions than the lattice vectors, which exhibit a stronger dependence on the initial atomic configuration.

Fig. 9(a,b) shows the most energetically stable HEA configuration, which exhibits periodicity in the *xy*-plane and provides two distinct surfaces for N incorporation. This configuration was subsequently selected as the precursor for HEN formation. The N incorporation process proceeds through two distinct stages. Initially, N atoms preferentially occupy interstitial positions between neighboring metal atoms, as shown in Fig. 9(c,d). As N incorporation progresses, the atoms gradually diffuse into the HEA lattice. The final configuration (Fig. 9e,f) confirms the presence of interstitial N distributed throughout the bulk structure. Upon completion of the N incorporation process, the 105-atom system contains a N mass fraction of approximately 10 wt%, while both exposed surfaces exhibit

a N density of approximately 20% relative to the number of atoms in the outermost surface layer.

The structural cohesion energy ($E_{\mathrm{cohe}}$) was also evaluated before and after nitrogen incorporation using eq. (3):

$$E_{cohe} = \frac{1}{N}\left(E_{HEA/HEN} - \sum_{i=1}^{n} N_i E_i\right) \quad (3)$$

Where $E_{\mathrm{HEA/HEN}}$ represents the total energy of the nanostructure (HEA and HEN), $n$ is the number of distinct atomic species, $N_i$ is the number of atoms for species ($i$), and $N$ represents the total number of atoms in the slab. The $E_{\mathrm{cohe}}$ were 6.86 and 6.73 eV for the HEA and HEN structures, respectively. The small energy difference ($\Delta E$) of 0.13 eV corresponds to an equivalent temperature of approximately 1160 K, estimated using the relationship, $T = \Delta E/k_B$, where $k_B$ is the Boltzmann constant ($8.617\times10^{-5}$ eV $K^{-1}$). The excellent agreement between the calculated equivalent temperature and the experimentally employed nitridation temperature further supports the thermodynamic feasibility of the N incorporation process.

Given the close agreement between the calculated equivalent temperature and the experimental synthesis conditions, the influence of N incorporation on the electronic and optical properties of the system was subsequently investigated. To elucidate the effect of N incorporation on the local electronic environment of the HEA, the spin-density distribution was calculated for both the pristine HEA and the HEN structures. As shown Fig. 9g,h, small regions of negative spin density are localized around the Cu and Ni atoms in the pristine HEA, whereas the remaining surface exhibits predominantly positive spin density. This distribution suggests a relatively low density of electronically active surface sites prior to N incorporation. In contrast, the HEN structure exhibits a substantial increase in the regions of negative spin density following N incorporation, indicating the generation of a greater number of potential active sites for surface interactions.

The influence of N on the optical properties was further examined by calculating the absorption coefficient (Fig. 9i). Compared with the pristine HEA, the HEN exhibits a modified optical absorption profile following N incorporation. This behavior is attributed to the coordination of nitrogen atoms with under-coordinated metallic sites on the HEA surface. By satisfying these coordination requirements, nitrogen effectively redistributes the electronic density, thereby modifying the electronic structure and consequently the optical response of the HEN system.

Following the identification of the most favorable adsorption sites for each antibiotic, total-energy minimizations were performed along the *z*-axis to obtain the equilibrium adsorption configurations on the HEN surface. The calculated binding energies for SME and TCL were −4.66 and −3.67 eV, respectively, with corresponding equilibrium adsorption distances of 2.73 and 2.66 Å. The large negative binding energies indicate thermodynamically favorable adsorption and strong molecule–surface interactions, exceeding the interaction strength typically associated with pure physisorption.

Fig. 10a,b illustrates the optimized adsorption geometries of SME and TCL on the HEN surface, showcasing their equilibrium configurations and local interfacial atomic arrangements. The two antibiotics adopt distinct orientations and adsorption configurations, reflecting differences in their molecular structures and functional groups. During structural optimization, no bond breakage or lattice disruption of the HEN was observed. Instead, only minor local rearrangements of the surface atoms occurred near the adsorption sites, promoting localized structural stabilization while preserving the crystallographic integrity of the substrate. Both SME and TCL underwent slight

translational and rotational adjustments during geometry optimization, enabling the molecules to attain energetically favorable adsorption configurations without significant distortion of their molecular backbones. These subtle structural relaxations enhance the interfacial contact between the antibiotics and the HEN surface, resulting in stable adsorption while maintaining the structural integrity of both the adsorbates and the substrate. The distinct adsorption orientations observed for SME and TCL are likely governed by differences in molecular size, functional-group distribution, and polarity. To further illustrate the adsorption process, the optimized structures and their dynamic evolution are provided in Supplementary Videos V2 and V3 for SME and TCL, respectively, which clearly demonstrate the molecular reorientation and stabilization leading to the equilibrium adsorption configurations.

These computational results corroborate the experimental observations by confirming the excellent structural stability of the HEN framework during molecular adsorption while revealing distinct adsorption configurations for the two antibiotics. Along with the LCTEM observations, the simulations provide atomistic insight into the catalyst–pollutant interactions that govern the photocatalytic degradation process.

### 3.6. Reusability of (MnFeCoNiCu)N NPs

For practical wastewater treatment applications, the long-term stability and reusability of a photocatalyst are as important as its initial catalytic activity. Accordingly, the cyclic stability of HEN-1:5 was evaluated through three consecutive photocatalytic degradation experiments under identical reaction conditions (Fig. 11a,b). After three successive cycles, HEN-1:5 retained high photocatalytic activity, achieving degradation efficiencies of 91% for SME and 89% for TCL, demonstrating its excellent operational stability and reusability. The slight decline in degradation efficiency is likely attributable to the partial accumulation of adsorbed intermediates or residual antibiotic molecules on the catalyst surface, along with a minor loss of catalytically active sites during repeated operation.

The environmental stability of HEN-1:5 was further assessed through metal-leaching analysis. As summarized in Table S3, the concentrations of all leached metal ions remained well below the permissible discharge limits for inland surface waters specified under the Environment (Protection) Act, 1986 (India). These findings demonstrate that the entropy-stabilized nitride framework effectively suppresses metal leaching while maintaining excellent photocatalytic performance, highlighting the environmental compatibility and long-term applicability of HEN-1:5 for wastewater treatment.

### 3.7. Performance in real water matrices

To assess the practical applicability of HEN-1:5 under environmentally relevant conditions, its photocatalytic performance toward the simultaneous degradation of SME and TCL was evaluated in different real water matrices (Fig. 11c). The physicochemical characteristics of the water samples are summarized in Table S4. Compared with the single-component system, the binary antibiotic mixture exhibited a slight reduction (4–6%) in degradation efficiency even in DI water, reflecting competition between SME and TCL molecules for the available active sites on the HEN-1:5 surface. The photocatalytic performance gradually declined with increasing water-matrix complexity. In tap water (TW), the degradation efficiencies decreased to 83% for SME and 81% for TCL, while further reductions to 79% and 75%, respectively, were observed in pond water (PW). This decline is primarily attributed to the presence of naturally occurring inorganic ions, particularly $Cl^-$ and $HCO_3^-$, which compete for reactive oxygen species (ROS) and suppress photocatalytic oxidation. A more pronounced decrease in degradation efficiency was observed in municipal wastewater (MWW), where only 57% of SME and 55% of TCL were degraded.

The inferior performance in MWW can be attributed to its higher turbidity, elevated concentrations of dissolved organic matter, inorganic ions ($HCO_3^-$, $HPO_4^{2-}$, and $SO_4^{2-}$), and total dissolved solids. These constituents reduce light penetration, scavenge reactive radicals, compete for catalytically active sites, and modify the solution chemistry, thereby collectively suppressing photocatalytic activity. To further examine this effect, the degradation experiments were repeated using treated municipal wastewater (T-MWW). As shown in Fig. 11c, the degradation efficiencies increased to 81% for SME and 75% for TCL, confirming that the removal of interfering constituents substantially improves photocatalytic performance. These findings demonstrate that although complex water matrices inevitably reduce photocatalytic efficiency, HEN-1:5 remains effective after appropriate pretreatment, highlighting its potential for practical wastewater treatment applications.

### 3.8. Phytotoxicity evaluation of photocatalytic degradation products

The phytotoxicity of the treated effluents was evaluated using *Vigna radiata* to determine whether photocatalytic degradation effectively reduced the toxicity of SME, TCL, and their transformation products. Seedlings were grown for 7 days in untreated antibiotic solutions (5 mg $L^{-1}$), photocatalytically treated solutions obtained using HEN-1:5, and deionized (DI) water as the control. As shown in Fig. S6, seedlings grown in DI water exhibited the greatest shoot and root lengths (11.8 and 4.1 cm, respectively), representing normal growth under non-toxic conditions. Exposure to untreated SME markedly inhibited seedling development, reducing the shoot and root lengths to 7.5 and 3.1 cm, respectively (Fig. S6b,f). Following photocatalytic treatment, however, the corresponding values increased to 11.2 and 3.8 cm (Fig. S6c,f), resulting in inhibition rates of only 5.1% (shoot) and 7.3% (root), with recovery efficiencies of 94.9% and 92.7%, respectively. These results demonstrate that photocatalytic degradation substantially reduced the phytotoxicity of SME and its degradation products. A similar trend was observed for TCL. Untreated TCL exhibited greater phytotoxicity than SME, reducing the shoot and root lengths to 6.4 and 2.8 cm, respectively (Fig. S6d,f). After photocatalytic treatment, seedling growth recovered considerably, with shoot and root lengths increasing to 9.5 and 3.5 cm, respectively (Fig. S6e,f). This corresponds to inhibition rates of 19.5% (shoot) and 14.6% (root), together with recovery efficiencies of 80.5% and 85.4%, confirming that HEN-1:5 effectively detoxifies TCL-containing solutions despite the higher intrinsic toxicity of the parent antibiotic. Overall, the marked reduction in phytotoxicity following photocatalytic treatment demonstrates that HEN-1:5 not only degrades SME and TCL but also effectively detoxifies their transformation products. The restoration of seedling growth to near-control levels highlight the environmental compatibility of the treated effluents and supports the potential application of HEN-1:5 for sustainable wastewater treatment and water reuse.

## 4. Conclusion

(MnFeCoNiCu)N HEN NPs were successfully synthesized through induction melting followed by nitridation and demonstrated excellent visible-light-driven photocatalytic activity toward the degradation of SME and TCL. Among the synthesized compositions, HEN-1:5 exhibited the highest activity, achieving degradation efficiencies of 96% for SME and 94% for TCL within 2 h, while maintaining excellent performance under repeated operation with negligible metal leaching. The catalyst also exhibited remarkable tolerance toward variations in operational parameters and retained high photocatalytic activity in real water matrices, highlighting its practical applicability for water treatment. Comprehensive experimental and theoretical investigations provided mechanistic insights into the photocatalytic process. Radical scavenging experiments established •OH radicals and photogenerated holes as the dominant reactive species responsible for antibiotic degradation. Temperature-dependent synchrotron ADXRD confirmed the exceptional structural stability of the entropy-stabilized nitride framework, while *in situ* LCTEM directly visualized the dynamic interactions between the catalyst and pollutant molecules under liquid-phase conditions. Complementary molecular simulations further revealed the structural robustness of the HEN framework during molecular adsorption and clarified the distinct adsorption configurations adopted by SME and TCL on the catalyst surface. Importantly, LC–MS analysis demonstrated the progressive degradation of both antibiotics into lower-molecular-weight intermediates, while phytotoxicity assays using *Vigna radiata* confirmed the effective detoxification of the treated effluents. These findings establish HENs as a promising class of robust and multifunctional photocatalysts that integrate high catalytic efficiency, structural durability, mechanistic transparency, and environmental compatibility, offering new opportunities for the development of advanced materials for sustainable water purification.

## References

[1] K. Saravanakumar, K. Yun, V. Maheskumar, Y. Yea, G. Jagan, C.M. Park, "Construction of novel $In_2S_3/Ti_3C_2$ MXene quantum dots/$SmFeO_3$ Z-scheme heterojunctions for efficient photocatalytic removal of sulfamethoxazole and 4-chlorophenol: Degradation pathways and mechanism insights," *Chemical Engineering Journal* 451 (2023): 138933. https://doi.org/10.1016/j.cej.2022.138933.

[2] J.S. Cedeño-Muñoz, S.A. Aransiola, K.V. Reddy, P. Ranjit, M.O. Victor-Ekwebelem, O.J. Oyedele, I.B. Pérez-Almeida, N.R. Maddela, J.M. Rodríguez-Díaz, "Antibiotic resistant bacteria and antibiotic resistance genes as contaminants of emerging concern: Occurrences, impacts, mitigations and future guidelines," *Science of The Total Environment* 952 (2024): 175906. https://doi.org/10.1016/j.scitotenv.2024.175906.

[3] Y. Chen, Y. Huang, Y. Chen, Y. Li, N. Sidikjan, N. Lin, Y. Li, X. Guo, G. Shen, M. Liu, "A systematic review of sources, occurrence, behavior and risks of global marine antibiotics," *Npj Emerging Contaminants* 1 (2025): 15. https://doi.org/10.1038/s44454-025-00015-z.

[4] S. Zhang, H. Zheng, P.G. Tratnyek, "Advanced redox processes for sustainable water treatment," *Nature Water* 1 (2023): 666–681. https://doi.org/10.1038/s44221-023-00098-1.

[5] S. Li, X. Wang, B. Xue, D. Feng, Y. Liu, W. Jiang, "Flower-like $Ag/Ag_2O/Bi_{12}O_{17}Cl_2$ heterojunction for photocatalytic removal of antibiotics: Synergetic effect of plasmonic effect and p–n heterojunction," *Journal of Materials Science & Technology* 246 (2026): 237–246. https://doi.org/10.1016/j.jmst.2024.12.088.

[6] S. Li, S. Yan, Z. Tong, X. Yong, X. Zhang, J. Zhou, "Assessment of photocatalytic activities of layered double hydroxide@petrochemical sludge biochar for sulfamethoxazole degradation," *Separation and Purification Technology* 355 (2025): 129732. https://doi.org/10.1016/j.seppur.2024.129732.

[7] W.-L. Hsu, C.-W. Tsai, A.-C. Yeh, J.-W. Yeh, "Clarifying the four core effects of high-entropy materials," *Nature Reviews Chemistry* 8 (2024): 471–485. https://doi.org/10.1038/s41570-024-00602-5.

[8] R. Huang, H. Zhao, Z. Chen, "High-entropy materials for photocatalysis," *Nano Materials Science* 8 (2026): 264–283. https://doi.org/10.1016/j.nanoms.2024.09.002.

[9] B.-F. Shan, J. Yang, X. Xiang, Z.-Y. Zhao, "High-entropy compounds for photo(electro)catalysis: diverse materials and applications," *Journal of Materials Chemistry A* 13 (2025): 12808–12827. https://doi.org/10.1039/D4TA09220A.

[10] B.R. Kc, B.P. Bastakoti, "Rational Design of High-Entropy Materials for Photo and Electrocatalytic Applications," *Small Structures* 6 (2025): 2500237. https://doi.org/10.1002/sstr.202500237.

[11] S. Schweidler, M. Botros, F. Strauss, Q. Wang, Y. Ma, L. Velasco, G. Cadilha Marques, A. Sarkar, C. Kübel, H. Hahn, J. Aghassi-Hagmann, T. Brezesinski, B. Breitung, "High-entropy materials for energy and electronic applications," *Nature Reviews Materials* 9 (2024): 266–281. https://doi.org/10.1038/s41578-024-00654-5.

[12] Q. Ding, Y. Zhang, X. Chen, X. Fu, D. Chen, S. Chen, L. Gu, F. Wei, H. Bei, Y. Gao, M. Wen, J. Li, Z. Zhang, T. Zhu, R.O. Ritchie, Q. Yu, "Tuning element distribution, structure and properties by composition in high-entropy alloys," *Nature 574* (2019): 223–227. https://doi.org/10.1038/s41586-019-1617-1.

[13] P. Sarker, T. Harrington, C. Toher, C. Oses, M. Samiee, J.-P. Maria, D.W. Brenner, K.S. Vecchio, S. Curtarolo, "High-entropy high-hardness metal carbides discovered by entropy descriptors," *Nature Communications* 9 (2018): 4980. https://doi.org/10.1038/s41467-018-07160-7.

[14] C.-Y. He, P. Zhao, X.-H. Gao, G. Liu, P.-Q. La, "Enhancing thermal robustness of a high-entropy nitride based solar selective absorber by the incorporation of Al element," *Materials Today Physics* 27 (2022): 100836. https://doi.org/10.1016/j.mtphys.2022.100836.

[15] E. Suhr, O.A. Krysiak, V. Strotkötter, F. Thelen, W. Schuhmann, A. Ludwig, "High-Throughput Exploration of Structural and Electrochemical Properties of the High-Entropy Nitride System (Ti–Co–Mo–Ta–W)N," *Advanced Engineering Materials* 25 (2023): 2300550. https://doi.org/10.1002/adem.202300550.

[16] A.M. Abdellah, K.E. Salem, L.-A. DiCecco, F. Ismail, A. Rakhsha, K. Grandfield, D. Higgins, "In Situ Transmission Electron Microscopy of Electrocatalyst Materials:

Proposed Workflows, Technical Advances, Challenges, and Lessons Learned,” *Small Methods* 9 (2025): 2400851. https://doi.org/10.1002/smtd.202400851.

[17] G. Ummethala, R. Jada, S. Dutta-Gupta, J. Park, A.H. Tavabi, S. Basak, R. Hooley, H. Sun, H.H. Pérez Garza, R.-A. Eichel, R.E. Dunin-Borkowski, S.R.K. Malladi, “Real-time visualisation of fast nanoscale processes during liquid reagent mixing by liquid cell transmission electron microscopy,” *Communications Chemistry* 8 (2025) 8. https://doi.org/10.1038/s42004-025-01407-3.

[18] Z. Manzoor, R. Karthik, M.A. Ferreira, D.S. Galvao, N.K. Mukhopadhyay, T.P. Yadav, P. Dadhwal, P.C. Saka, C.F. Woellner, S. Chowdhury, C.S. Tiwary, “Radio frequency-induced catalysis using multi-component two-dimensional quasicrystals for effective sulfamethoxazole removal from water,” *Applied Catalysis B: Environment and Energy* 383 (2026): 126062. https://doi.org/10.1016/j.apcatb.2025.126062.

[19] A. García, N. Papior, A. Akhtar, E. Artacho, V. Blum, E. Bosoni, P. Brandimarte, M. Brandbyge, J.I. Cerdá, F. Corsetti, R. Cuadrado, V. Dikan, J. Ferrer, J. Gale, P. García-Fernández, V.M. García-Suárez, S. García, G. Huhs, S. Illera, R. Korytár, P. Koval, I. Lebedeva, L. Lin, P. López-Tarifa, S.G. Mayo, S. Mohr, P. Ordejón, A. Postnikov, Y. Pouillon, M. Pruneda, R. Robles, D. Sánchez-Portal, J.M. Soler, R. Ullah, V.W. Yu, J. Junquera, “Siesta: Recent developments and applications,” *The Journal of Chemical Physics* 152 (2020) 204108. https://doi.org/10.1063/5.0005077.

[20] J.M. Soler, E. Artacho, J.D. Gale, A. García, J. Junquera, P. Ordejón, D. Sánchez-Portal, “The SIESTA method for ab initio order-N materials simulation,” *Journal of Physics Condensed Matter* 14 (2002): 2745. https://doi.org/10.1088/0953-8984/14/11/302.

[21] J.P. Perdew, K. Burke, M. Ernzerhof, “Generalized Gradient Approximation Made Simple,” *Physical Review Letters* 77 (1996): 3865–3868. https://doi.org/10.1103/PhysRevLett.77.3865.

[22] S.L. Dudarev, G.A. Botton, S.Y. Savrasov, C.J. Humphreys, A.P. Sutton, "Electron-energy-loss spectra and the structural stability of nickel oxide: An LSDA+U study," *Physical Review B* 57 (1998) 1505–1509. https://doi.org/10.1103/PhysRevB.57.1505.

[23] H.J. Monkhorst, "Special points for Brillouin-zone integrations," *Physical Review B* 13 (1976): 5188–5192. https://doi.org/10.1103/PhysRevB.13.5188.

[24] A. Hjorth Larsen, J. Jørgen Mortensen, J. Blomqvist, I.E. Castelli, R. Christensen, M. Dułak, J. Friis, M.N. Groves, B. Hammer, C. Hargus, E.D. Hermes, P.C. Jennings, P. Bjerre Jensen, J. Kermode, J.R. Kitchin, E. Leonhard Kolsbjerg, J. Kubal, K. Kaasbjerg, S. Lysgaard, J. Bergmann Maronsson, T. Maxson, T. Olsen, L. Pastewka, A. Peterson, C. Rostgaard, J. Schiøtz, O. Schütt, M. Strange, K.S. Thygesen, T. Vegge, L. Vilhelmsen, M. Walter, Z. Zeng, K.W. Jacobsen, "The atomic simulation environment—a Python library for working with atoms," *Journal of Physics Condened Matter* 29 (2017): 273002. https://doi.org/10.1088/1361-648X/aa680e.

[25] R.M. Tromer, L.D. Machado, C.F. Woellner, D.S. Galvao, "Thiophene-Tetrathia-Annulene monolayer (TTA-2D): A new 2D semiconductor material with indirect bandgap," *Physica E: Low-Dimensional Systems and Nanostructures* 129 (2021): 114586. https://doi.org/10.1016/j.physe.2020.114586.

[26] S. Grimme, C. Bannwarth, P. Shushkov, "A Robust and Accurate Tight-Binding Quantum Chemical Method for Structures, Vibrational Frequencies, and Noncovalent Interactions of Large Molecular Systems Parametrized for All spd-Block Elements (Z = 1 –86)," *Journal of Chemical Theory and Computation* 13 (2017): 1989–2009. https://doi.org/10.1021/acs.jctc.7b00118.

[27] B. Hourahine, B. Aradi, V. Blum, F. Bonafé, A. Buccheri, C. Camacho, C. Cevallos, M.Y. Deshaye, T. Dumitrică, A. Dominguez, S. Ehlert, M. Elstner, T. van der Heide, J. Hermann, S. Irle, J.J. Kranz, C. Köhler, T. Kowalczyk, T. Kubař, I.S. Lee, V. Lutsker, R.J.

Maurer, S.K. Min, I. Mitchell, C. Negre, T.A. Niehaus, A.M.N. Niklasson, A.J. Page, A. Pecchia, G. Penazzi, M.P. Persson, J. Řezáč, C.G. Sánchez, M. Sternberg, M. Stöhr, F. Stuckenberg, A. Tkatchenko, V.W. -z. Yu, T. Frauenheim, “DFTB+, a software package for efficient approximate density functional theory based atomistic simulations,” *The Journal of Chemical Physics* 152 (2020): 124101. https://doi.org/10.1063/1.5143190.

[28] J. Hei, N. Wang, R. Jing, X. Chen, X. Yin, J. Li, P. Zuo, Y. Yin, L. Cui, “High-Entropy Nitrides as Superior Electrocatalysts: Unveiling the Role of Entropy in Enhanced Performance,” *Chemistry–A European Journal* 31 (2025): e202500039. https://doi.org/10.1002/chem.202500039.

[29] H. Qiu, S. Peng, Y. Zou, X. Cui, Y. Zhao, X. Wang, D. Sun, “Effect of nitrogen content on mechanical properties and microstructure of (TiZrNbCrSi)Nx high-entropy nitride coating,” *Journal of Alloys and Compounds* 1022 (2025): 179822. https://doi.org/10.1016/j.jallcom.2025.179822.

[30] R. Wang, J. Jiao, D. Liu, Y. He, Y. Yang, D. Sun, H. Pan, F. Fang, R. Wu, “High-Entropy Metal Nitride Embedded in Concave Porous Carbon Enabling Polysulfide Conversion in Lithium–Sulfur Batteries,” *Small* 20 (2024): 2405148. https://doi.org/10.1002/smll.202405148.

[31] L. Yang, W. Chen, C. Sheng, H. Wu, N. Mao, H. Zhang, “Fe/N-codoped carbocatalysts loaded on carbon cloth (CC) for activating peroxymonosulfate (PMS) to degrade methyl orange dyes,” *Applied Surface Science* 549 (2021): 149300. https://doi.org/10.1016/j.apsusc.2021.149300.

[32] K.S. Anu, K.A. Vishnumurthy, A. Mahesh, K. Natarajan, “Carbon fiber-reinforced, activated carbon-embedded copper oxide nanoparticles/epoxy hybrid composites for EMI shielding in aircraft applications,” *Polymer Bulletin* 81 (2024): 8723–8750. https://doi.org/10.1007/s00289-023-05112-w.

[33] A. Aygun, E. Ozveren, E. Halvaci, D. Ikballi, R.N.E. Tiri, C. Catal, M. Bekmezci, A. Ozengul, I. Kaynak, F. Sen, "The performance of a very sensitive glucose sensor developed with copper nanostructure-supported nitrogen-doped carbon quantum dots," *RSC Advances* 14 (2024): 34964–34970. https://doi.org/10.1039/D4RA06566B.

[34] A. Babu, A.S. Dsouza, A. Viswanathan, A.N. Shetty, "Single-Step Synthesis and Characterizations of Polyaniline/$MnO_2$ Nanocomposites for Its Durable Supercapacitive Application," *Energy Technology* 14 (2026): e70505. https://doi.org/10.1002/ente.70505.

[35] L.R.L. Mary, A.J. Pearl, "Comparative study on the photocatalytic efficiency of manganese dioxide nanoparticles synthesized by green and chemical methods," *Journal of Materials Science: Materials in Electronics* 36 (2025) 1425. https://doi.org/10.1007/s10854-025-15386-7.

[36] Z. Abasali karaj abad, A. Nemati, A. Malek Khachatourian, M. Golmohammad, "Synthesis and characterization of rGO/$Fe_2O_3$ nanocomposite as an efficient supercapacitor electrode material," *Journal of Materials Science: Materials in Electronics* 31 (2020): 14998–15005. https://doi.org/10.1007/s10854-020-04062-7.

[37] Ravina, G. Srivastava, S. Dalela, S. Kumar, M. Nasit, J. Singh, M.A. Ahmad, P.A. Alvi, "Study of structural, optical, surface and electrochemical properties of Co3O4 nanoparticles for energy storage applications," *Interactions* 245 (2024): 85. https://doi.org/10.1007/s10751-024-01932-y.

[38] S. Das, S. Kumar, S. Sarkar, D. Pradhan, C.S. Tiwary, S. Chowdhury, "High entropy spinel oxide nanoparticles for visible light-assisted photocatalytic degradation of binary mixture of antibiotic pollutants in different water matrixes," *Journal of Materials Chemistry A* 12 (2024): 16815–16830. https://doi.org/10.1039/D4TA02294G.

[39] Y. Liu, J. Zhang, W. Wang, L. Cao, B. Dong, "Two-Phase Colloidal Synthesis of Amorphous Iron-Doped Manganese Phosphate Hollow Nanospheres for Efficient Water

Oxidation," *Advanced Sustainable Systems* 4 (2020): 2000128. https://doi.org/10.1002/adsu.202000128.

[40] E. Samuel, B. Joshi, Y. Kim, C. Park, A. Aldalbahi, M. El-Newehy, H.-S. Lee, S.S. Yoon, "Cotton fabric decorated with manganese oxide nanorods as a supercapacitive flexible electrode for wearable electronics," *Applied Surface Science* 568 (2021): 150968. https://doi.org/10.1016/j.apsusc.2021.150968.

[41] S. Das, M. Sain, Z. Manzoor, S. Kumar, S. Sarkar, D. Pradhan, C.S. Tiwary, S. Chowdhury, "High entropy oxide nanoparticles for simultaneous photocatalytic degradation of antibiotics, analgesics, and antiepileptics in real water systems under continuous flow mode," *Journal of Environmental Management* 388 (2025): 125975. https://doi.org/10.1016/j.jenvman.2025.125975.

[42] K. Peng, L. Liu, N. Bhuvanendran, F. Qiao, G. Lei, S. Youn Lee, Q. Xu, H. Su, "Effective regulation on catalytic performance of nickel–iron–vanadium layered double hydroxide for urea oxidation via sulfur incorporation," *Materials Advances* 4 (2023) 1354–1362. https://doi.org/10.1039/D2MA01066F.

[43] P. Gao, C. Yue, J. Zhang, J. Bao, H. Wang, Q. Chen, Y. Jiang, S. Huang, Z. Hu, J. Zhang, "Construction of unique NiCoP/FeNiCoP hollow heterostructured ellipsoids with modulated electronic structure for enhanced overall water splitting," *Journal of Colloid and Interface Science* 666 (2024): 403–415. https://doi.org/10.1016/j.jcis.2024.03.198.

[44] L. Zhou, J. Yu, D. Wei, C. Shi, H. Jin, H. Li, E. Zhu, M. Xu, "Electronic structure engineering of Fe-NC catalysts by proximity electronic effect of S ligand for enhanced oxygen electrocatalysis," *Carbon Neutral Systems* 1 (2025): 18. https://doi.org/10.1007/s44438-025-00019-7.

[45] P. Thangasamy, R. He, H. Randriamahazaka, X. Chen, Y. Zhang, H. Luo, H. Wang, M. Zhou, "Collectively exhaustive electrochemical hydrogen evolution reaction of

polymorphic cobalt selenides derived from organic surfactants modified Co-MOFs," *Applied Catalysis B: Environmental* 325 (2023): 122367. https://doi.org/10.1016/j.apcatb.2023.122367.

[46] J.A. Torres-Ochoa, D. Cabrera-German, O. Cortazar-Martinez, M. Bravo-Sanchez, G. Gomez-Sosa, A. Herrera-Gomez, "Peak-fitting of Cu 2*p* photoemission spectra in $Cu^0$, $Cu^{1+}$, and $Cu^{2+}$ oxides: A method for discriminating $Cu^0$ from $Cu^{1+}$," *Applied Surface Science* 622 (2023) 156960. https://doi.org/10.1016/j.apsusc.2023.156960.

[47] L. Tang, L. Zhang, K. Ni, W. Xu, C. Kou, Q. Ren, R. Du, B. Zhou, B. Zhang, "2D Polymeric N-Heterocyclic Carbene Copper Microsheets Enable Tandem Electroreduction of Nitrate to Ammonia," *Advanced Functional Materials* 36 (2026): e23772. https://doi.org/10.1002/adfm.202523772.

[48] A. Ferlazzo, S. Bonforte, F. Florio, S. Petralia, L. Sorace, B. Muzzi, A. Caneschi, A. Gulino, "Photochemical eco-friendly synthesis of photothermal and emissive copper nanoclusters in water: towards sustainable nanomaterials," *Materials Advances* 5 (2024): 8034–8041. https://doi.org/10.1039/D4MA00401A.

[49] D.Yu. Osadchii, A.I. Olivos-Suarez, A.V. Bavykina, J. Gascon, "Revisiting Nitrogen Species in Covalent Triazine Frameworks," *Langmuir* 33 (2017): 14278–14285. https://doi.org/10.1021/acs.langmuir.7b02929.

[50] B.-B. Zhang, J. Ming, H.-Q. Li, X.-N. Song, C.-K. Wang, W. Hua, Y. Ma, "Structural characterization of nitrogen-doped γ-graphynes by computational X-ray spectroscopy," *Carbon* 214 (2023): 118301. https://doi.org/10.1016/j.carbon.2023.118301.

[51] D. Dey, S. Chowdhury, R. Sen, "Facile fabrication of a Z-scheme PVA/gelatin based $CeO_2$/g-$C_3N_4$ heterojunction aerogel for enhanced visible light mediated photocatalytic degradation of psychoactive drug in aqueous phase," *Next Materials* 8 (2025): 100533. https://doi.org/10.1016/j.nxmate.2025.100533.

[52] M. Machreki, T. Chouki, G. Tyuliev, M. Fanetti, M. Valant, D. Arčon, M. Pregelj, S. Emin, "The Role of Lattice Defects on the Optical Properties of $TiO_2$ Nanotube Arrays for Synergistic Water Splitting," *ACS Omega* 8 (2023): 33255–33265. https://doi.org/10.1021/acsomega.3c00965.

[53] W. Li, Y. Sun, L. Ye, W. Han, F. Chen, J. Zhang, T. Zhao, "Preparation of high entropy nitride ceramic nanofibers from liquid precursor for $CO_2$ photocatalytic reduction," *Journal of the American Ceramic Society* 105 (2022): 3729–3734. https://doi.org/10.1111/jace.18384.

[54] X. Lu, H. Li, J. Liang, D. Huo, C. Liu, M. Hu, Y. Yang, "A review of high-entropy nitrides: Preparation methods, properties, and applications," *Materials Today Communications* 47 (2025): 113092. https://doi.org/10.1016/j.mtcomm.2025.113092.

[55] M.V.L. Gnanaguru, Mu. Naushad, T. Tatarchuk, M.M. Ghangrekar, S. Chowdhury, "One-step calcination synthesis of 2D/2D g-$C_3N_4$/$WS_2$ van der Waals heterojunction for visible light-induced photocatalytic degradation of pharmaceutical pollutants," *Environmental Science and Pollution Research* 30 (2023): 78537–78553. https://doi.org/10.1007/s11356-023-27714-7.

[56] L. Luo, Z. Sun, Y. Chen, H. Zhang, Y. Sun, D. Lu, J. Ma, "Catalytic ozonation of sulfamethoxazole using low-cost natural silicate ore supported $Fe_2O_3$: influencing factors, reaction mechanisms and degradation pathways," *RSC Advances* 13 (2023): 1906–1913. https://doi.org/10.1039/D2RA06714E.

[57] X. Fu, Y. Lin, C. Yang, S. Wu, Y. Wang, X. Li, "Peroxymonosulfate activation via CoP nanoparticles confined in nitrogen-doped porous carbon for enhanced degradation of sulfamethoxazole in wastewater with high salinity," *Journal of Environmental Chemical Engineering* 10 (2022): 107734. https://doi.org/10.1016/j.jece.2022.107734.

[58] S.S. Khan, J.P. Steffy, A.T. Alfagham, A.M. Elgorban, “Engineering oxygen vacancies rich nano-heterojunction for enhanced visible-light-driven photo-Fenton degradation of sulfamethoxazole,” *Journal of Water Process Engineering* 72 (2025): 107542. https://doi.org/10.1016/j.jwpe.2025.107542.

[59] Y. Liu, G. Sun, Y. Cao, X. Li, Z. Li, Z. Chen, “Synergistic elimination of antibiotic resistance genes and tetracycline antibiotics in wastewater via a Z-scheme $Bi_2WO_6$/g-$C_3N_4$ heterojunction: degradation pathways and mechanism,” *RSC Advances* 16 (2026): 16424–16441. https://doi.org/10.1039/D5RA09524G.

[60] M. Sharma, M.K. Mandal, S. Pandey, R. Kumar, K.K. Dubey, “Visible-Light-Driven Photocatalytic Degradation of Tetracycline Using Heterostructured $Cu_2O$–$TiO_2$ Nanotubes, Kinetics, and Toxicity Evaluation of Degraded Products on Cell Lines,” *ACS Omega 7* (2022): 33572–33586. https://doi.org/10.1021/acsomega.2c04576.

[61] G. Yang, Y. Liang, Z. Xiong, J. Yang, K. Wang, Z. Zeng, “Molten salt-assisted synthesis of $Ce_4O_7$/$Bi_4MoO_9$ heterojunction photocatalysts for Photo-Fenton degradation of tetracycline: Enhanced mechanism, degradation pathway and products toxicity assessment,” *Chemical Engineering Journal* 425 (2021): 130689. https://doi.org/10.1016/j.cej.2021.130689.

[62] C. Li, S. Lin, W. Zhang, K. Qi, “Enhanced photocatalytic degradation of tetracycline by Mn, S co-doped $TiO_2$ under visible-light irradiation,” *Journal of Photochemistry and Photobiology A: Chemistry* 467 (2025): 116423. https://doi.org/10.1016/j.jphotochem.2025.116423.

[63] Y. Sun, M. Peng, Q. Zhang, X. Zhang, X. Xu, E. Du, “Aging of microplastics and its photo-catalytic degradation of coexisting tetracycline,” *Separation and Purification Technology* 355 (2025): 129605. https://doi.org/10.1016/j.seppur.2024.129605.

[64] S. Liu, N. Wang, Y. Li, D. Liang, W. He, N. Lin, J. Li, Y. Feng, "Magnetic biochar-mediated iron reduction synergizes with microbial co-metabolism for enhanced adsorption and in-situ degradation of sulfamethoxazole," *Chemical Engineering Journal* 523 (2025): 168620. https://doi.org/10.1016/j.cej.2025.168620.

[65] P. Kalimuthu, G. Kalaiyarasan, J. Jung, D. Tiwari, "Upcycled PET carbon dots build an S-scheme $Fe_2O_3$ heterojunction photocatalyst for visible-light degradation of tetracycline," *Separation and Purification Technology* 392 (2026): 137122. https://doi.org/10.1016/j.seppur.2026.137122.

[66] Y. Wang, J. Liu, B. Yan, Y. Wang, X. Li, M. Fang, G. Li, P. Rao, Y. Liu, "Engineering vacancy Rich MOFs on SiC ceramic membranes for robust tetracycline removal across a wide pH range," *Journal of Water Process Engineering* 85 (2026): 109739. https://doi.org/10.1016/j.jwpe.2026.109739.

[67] H. Liu, J. Cheng, H. Sun, X. Sun, "Boosting visible-light photocatalysis of tetracycline through $Eu^{3+}$ doping in hydrothermally synthesized $Bi_2WO_6$ composites," *Journal of Alloys and Compounds* 1032 (2025): 181116. https://doi.org/10.1016/j.jallcom.2025.181116.

[68] Q. Fu, L. Shi, Y. Tao, M. Tang, S. Hou, L. Ouyang, H. Zhou, J. Xu, "Design of FeN co-doped porous carbon catalysts via black fungus biomass pyrolysis: Mechanism and application in tetracycline removal," *Separation and Purification Technology* 390 (2026): 136896. https://doi.org/10.1016/j.seppur.2026.136896.

[69] Z. Yang, X. Liu, X. Hao, H. Yin, Y. Li, G. Zhang, "Bicarbonate-triggered in-situ epitaxial construction of lattice matched $NaBiO_3/Bi_2O_2CO_3$ Z-scheme heterojunction for enhanced photocatalytic degradation of organic pollutants: Boosted carrier separation and mechanism insight," *Applied Catalysis B: Environment and Energy* 386 (2026): 126389. https://doi.org/10.1016/j.apcatb.2026.126389.

[70] H. Kamaruddin, Z. Jianghong, L. Yu, W. Yuefan, H. Yizhong, “A review of noble metal-free high entropy alloys for water splitting applications,” *Journal of Materials Chemistry A* 12 (2024): 9933–9961. https://doi.org/10.1039/D4TA00690A.

[71] F. Wang, S. Chen, Z. Tong, X. Liu, H. Zhou, Y. He, G. Shu, Y. Jin, W. Zheng, “Electronegativity-Modulated PtFeCoNiCu High-Entropy Alloy Catalysts for Efficient HER and OER,” *ACS Applied Materials & Interfaces* 17 (2025): 66617–66629. https://doi.org/10.1021/acsami.5c17409.

# Figures

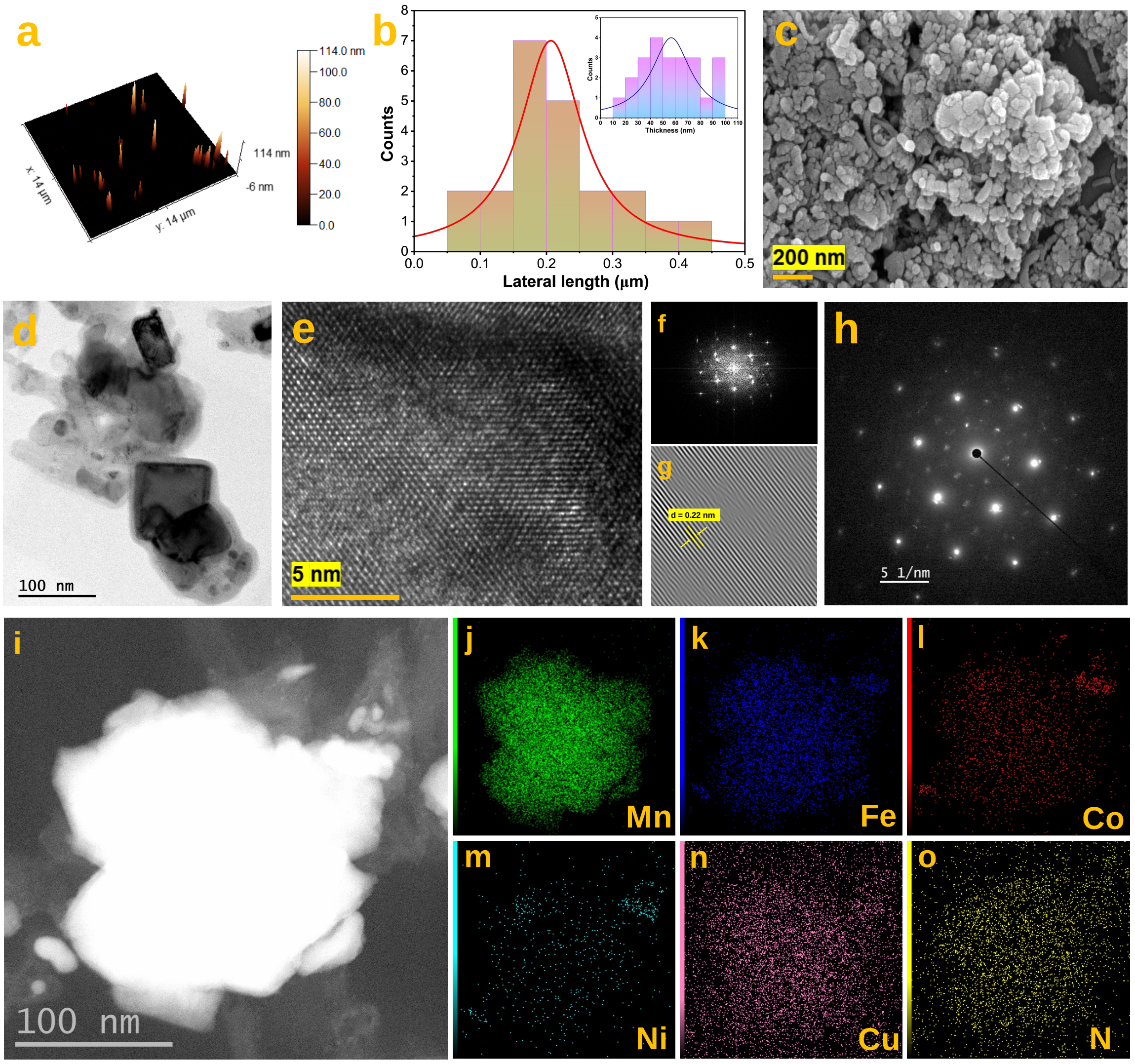

**Fig. 1:** (a) AFM image and (b) the corresponding lateral dimension distribution and the inset thickness profile of (MnFeCoNiCu)N HEN-1:5 NPs. (c) FESEM image, (d) bright-field TEM image, (f) HRTEM image, and (f,g) the FFT and inverse FFT patterns of (MnFeCoNiCu)N HEN-1:5 NPs. (h) SAED pattern and (i-o) STEM–EDS elemental mappings of (MnFeCoNiCu)N HEN-1:5 NPs.

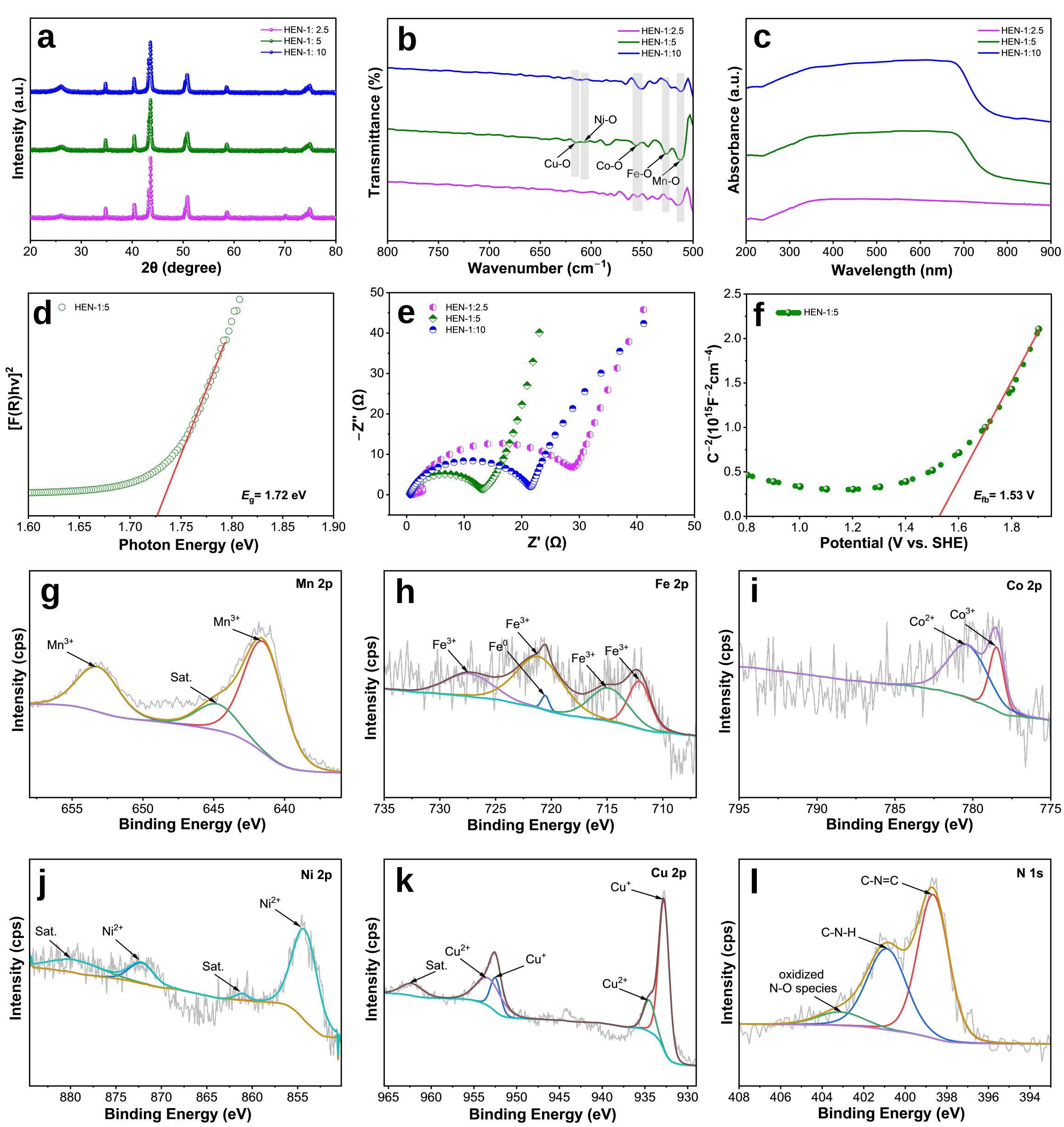

**Fig. 2:** (a) XRD pattern, (b) FTIR spectra, and (c) UV–visible diffuse reflectance spectra of (MnFeCoNiCu)N HEN NPs. (d) Kubelka–Munk plot of (MnFeCoNiCu)N HEN-1:5 NPs for estimating the bandgap energy. (e) Electrochemical impedance spectrum of the (MnFeCoNiCu)N HENs NPs. (f) Mott-Schottky analysis of (MnFeCoNiCu)N HEN-1:5 NPs. High-resolution deconvoluted XPS spectra of (g) Mn 2p, (h) Fe 2p, (i) Co 2p, (j) Ni 2p, (k) Cu 2p, and (l) N 1s of the (MnFeCoNiCu)N HEN-1:5 NPs.

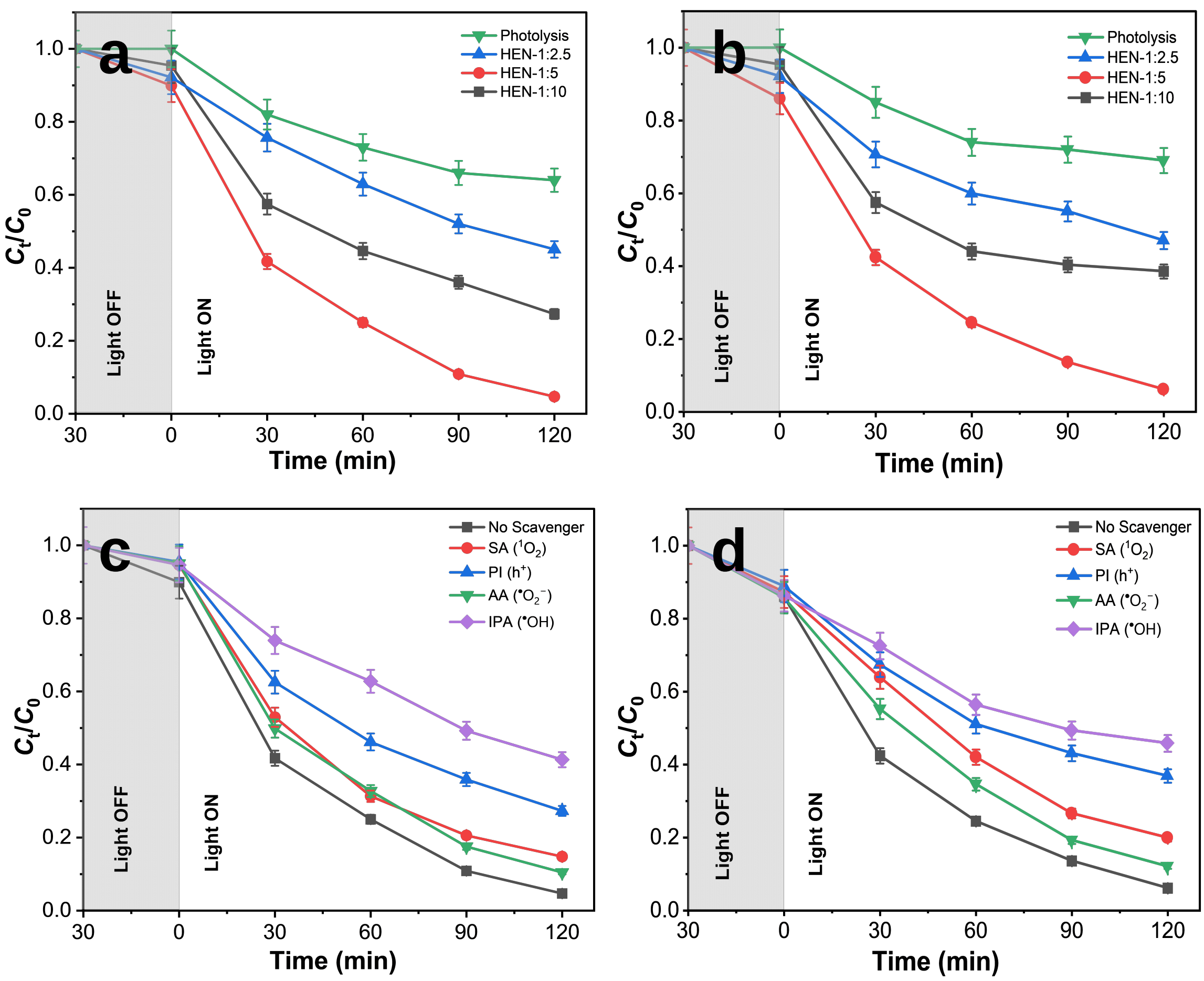

**Fig. 3:** (a) Visible light-induced photocatalytic degradation of (a) SME and (b) TCL over (MnFeCoNiCu)N HEN-NPs (experimental conditions: initial antibiotic concentration = 5 mg $L^{-1}$; photocatalyst dose = 0.50 g $L^{-1}$; temperature = 25 °C). Photocatalytic degradation of (c) SME and (d) TCL with and without ROS scavengers and hole quencher over (MnFeCoNiCu)N HEN-1:5 NPs (experimental conditions: initial antibiotic concentration = 5 mg $L^{-1}$; photocatalyst dose = 0.50 g $L^{-1}$; temperature = 25 °C).

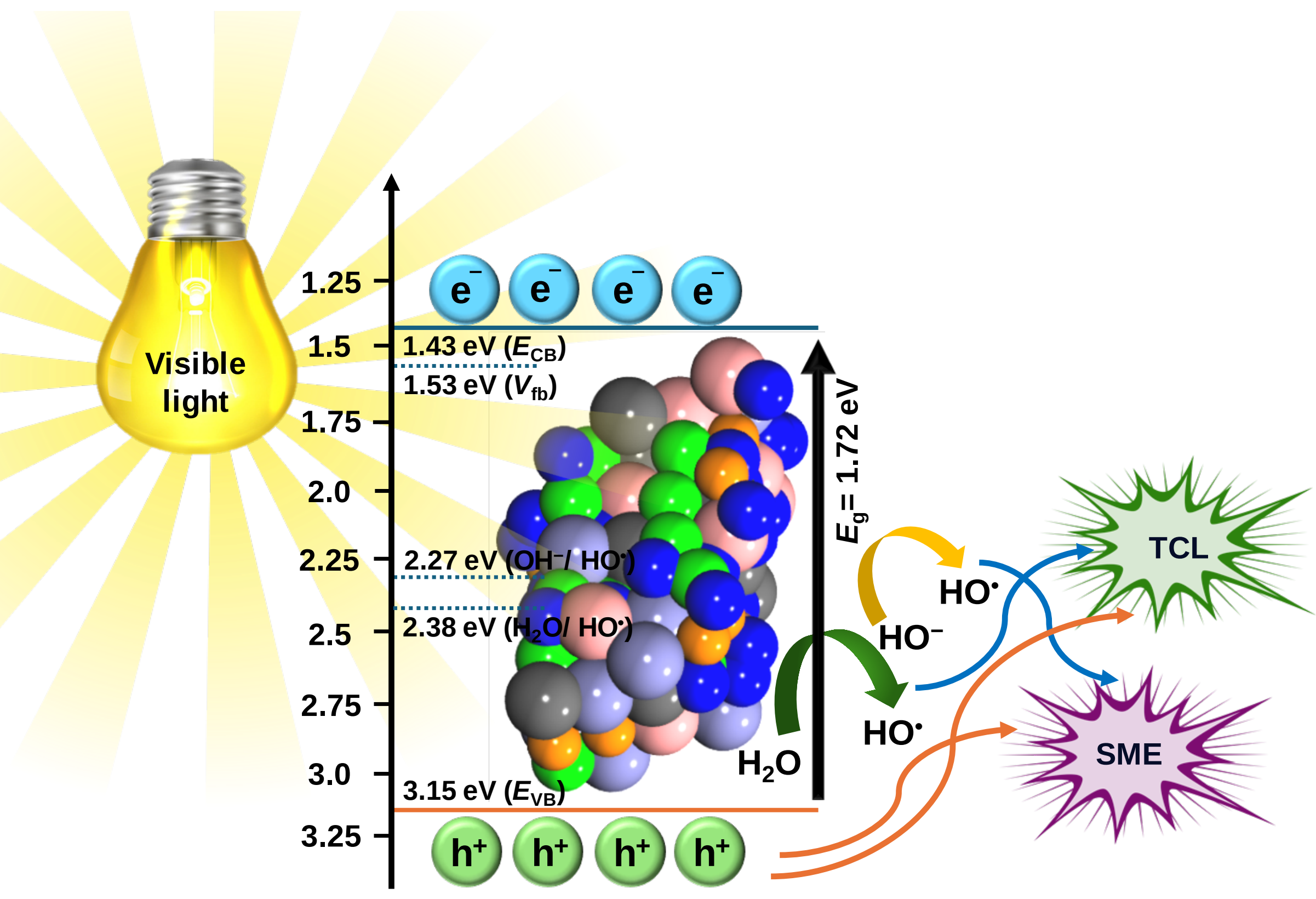

**Fig. 4:** Schematic of the plausible photocatalytic degradation mechanism of SME and TCL over (MnFeCoNiCu)N HEN-1:5 NPs under visible-light irradiation.

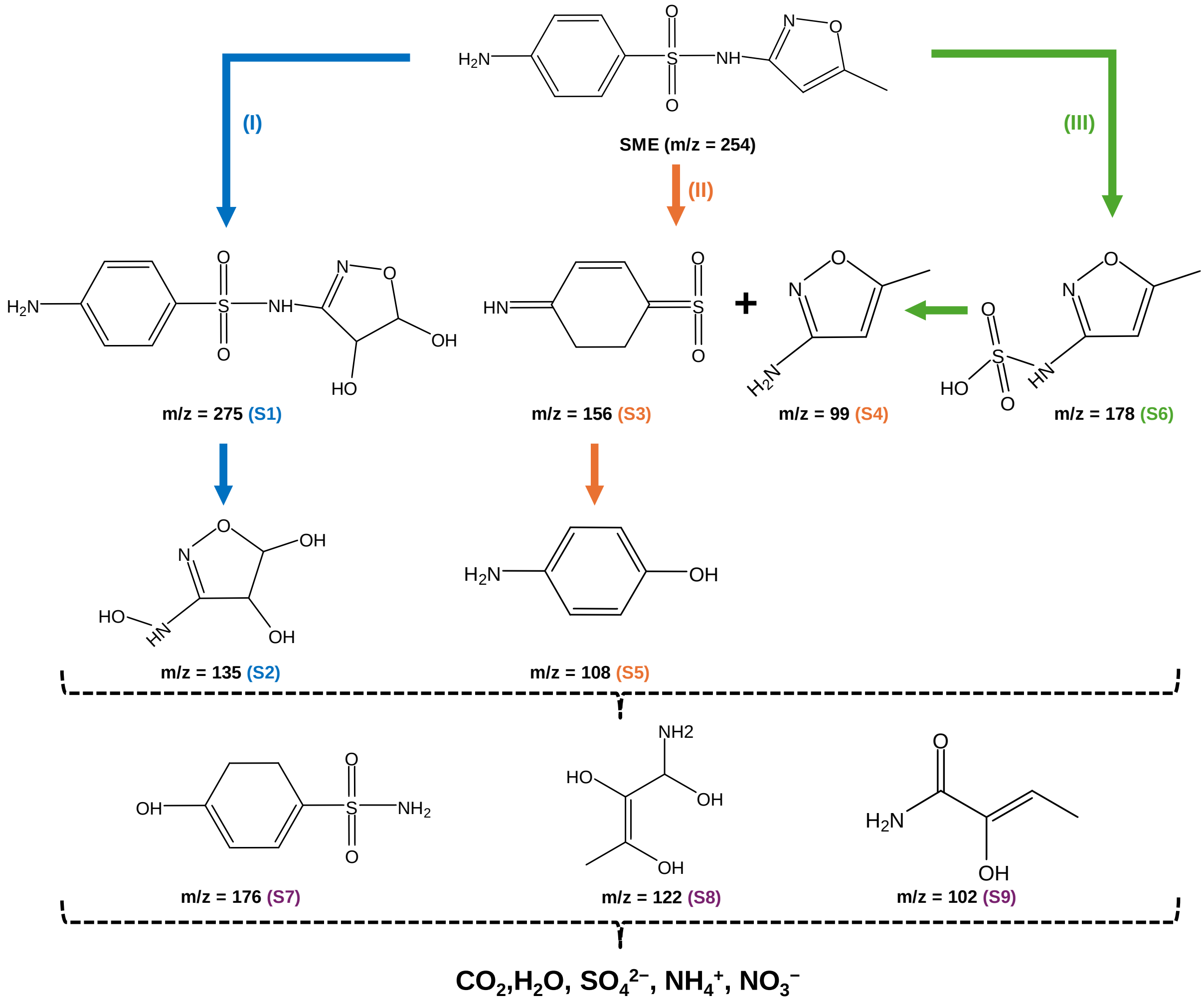

**Fig. 5:** Proposed pathway for the photocatalytic degradation of SME over (MnFeCoNiCu)N HEN-1:5 NPs under visible-light irradiation.

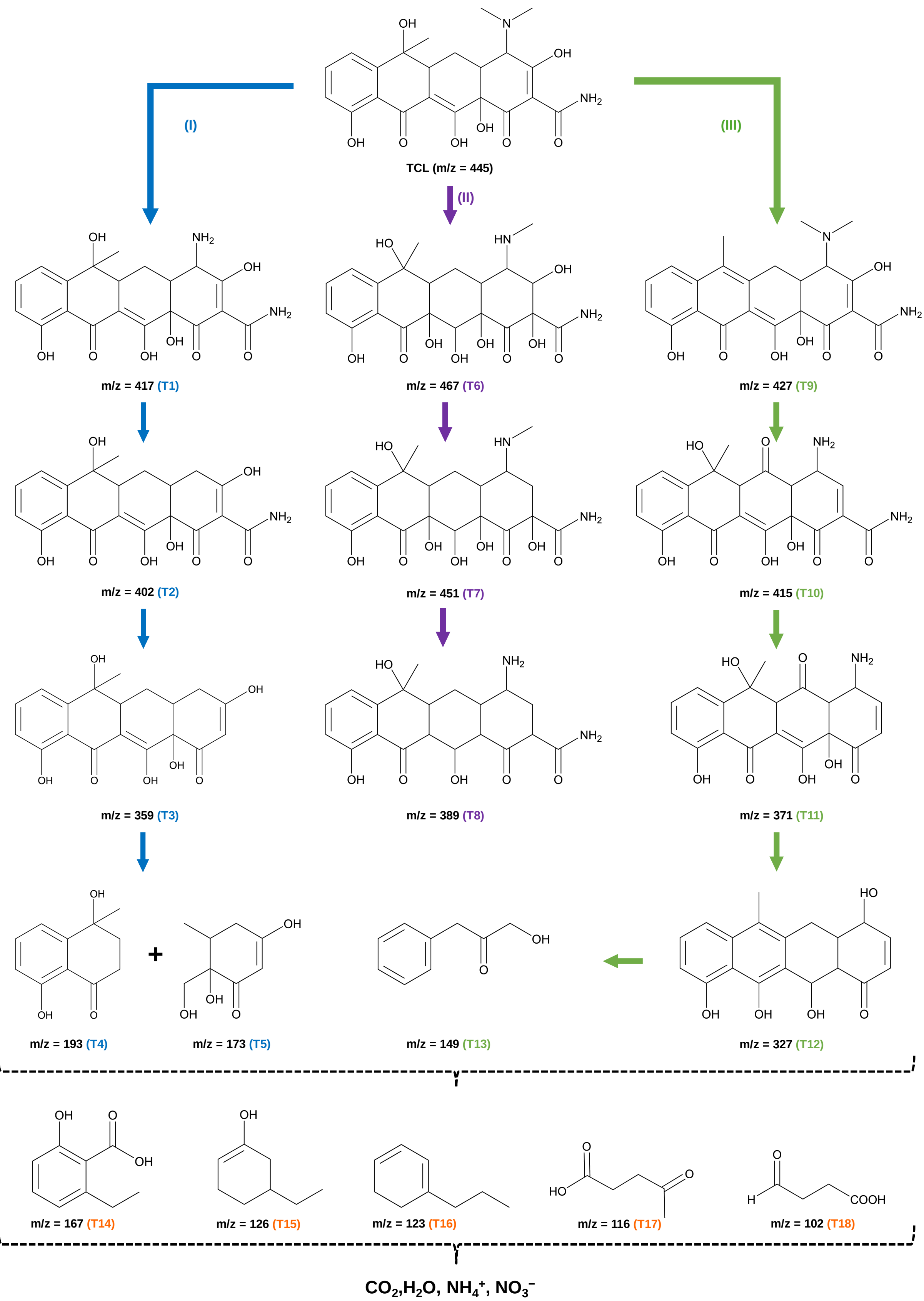

**Fig. 6:** Proposed pathway for the photocatalytic degradation of TCL over (MnFeCoNiCu)N HEN-1:5 NPs under visible-light irradiation.

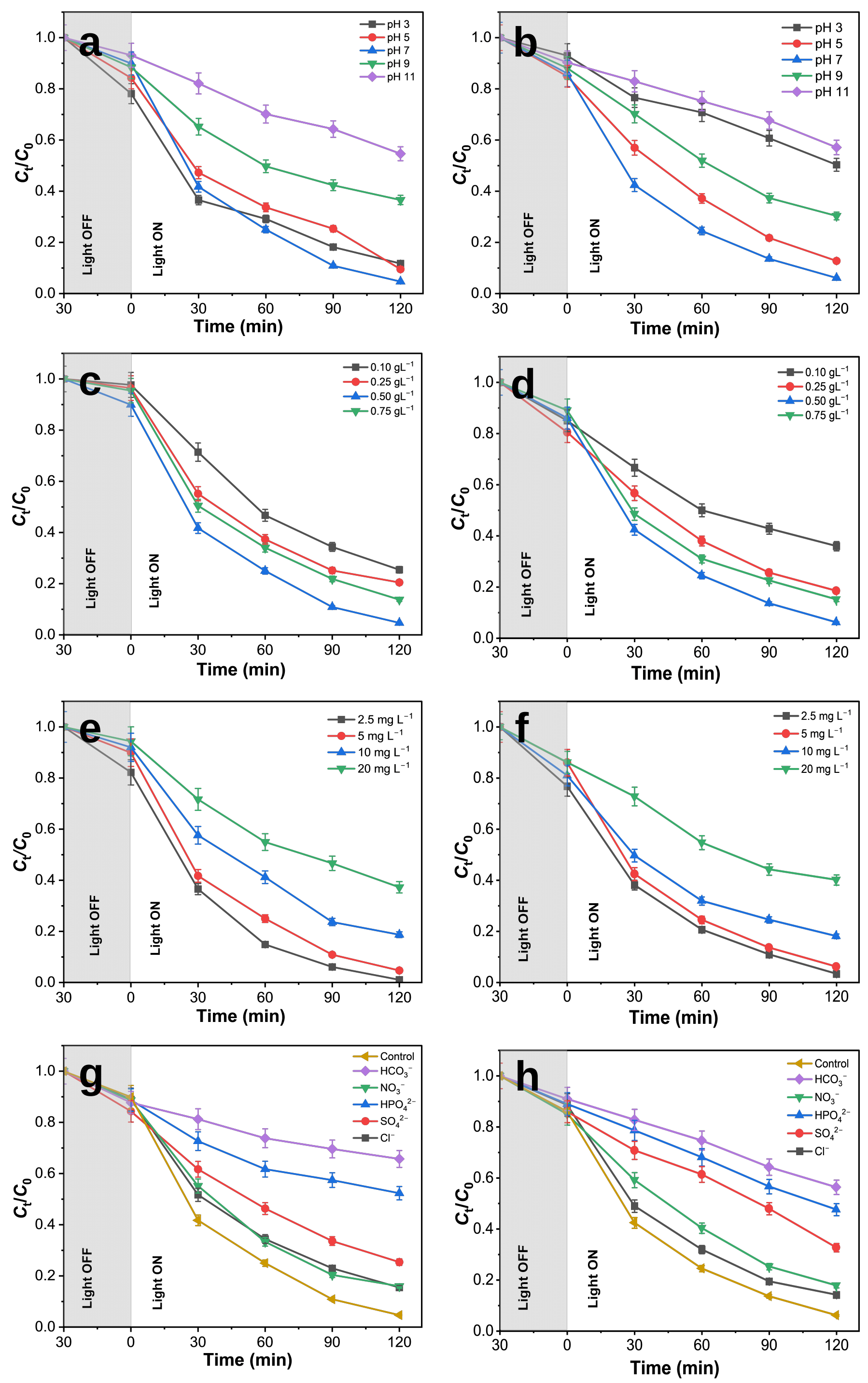

**Fig. 7:** Effect of the initial solution pH on the photocatalytic degradation of (a) SME and (b) TCL by (MnFeCoNiCu)N HEN-1:5 NPs under visible-light irradiation (experimental conditions: initial antibiotic concentration = 5 mg $L^{-1}$; photocatalyst dose = 0.50 g $L^{-1}$; temperature = 25 °C). Effect of the catalyst dose on the photocatalytic degradation of (c) SME and (d) TCL by (MnFeCoNiCu)N HEN-1:5 NPs under visible-light irradiation (experimental conditions: initial antibiotic concentration = 5 mg $L^{-1}$; temperature = 25 °C). Effect of the initial

antibiotic concentration on the photocatalytic degradation of (e) SME and (f) TCL by (MnFeCoNiCu)N HEN-1:5 NPs under visible-light irradiation (experimental conditions: photocatalyst dose = 0.50 g $L^{-1}$; temperature = 25 °C). Effect of inorganic anions on the photocatalytic degradation of (g) SME and (h) TCL by (MnFeCoNiCu)N HEN-1:5 NPs under visible-light irradiation (experimental conditions: initial antibiotic concentration = 5 mg $L^{-1}$; photocatalyst dose = 0.50 g $L^{-1}$; temperature = 25 °C).

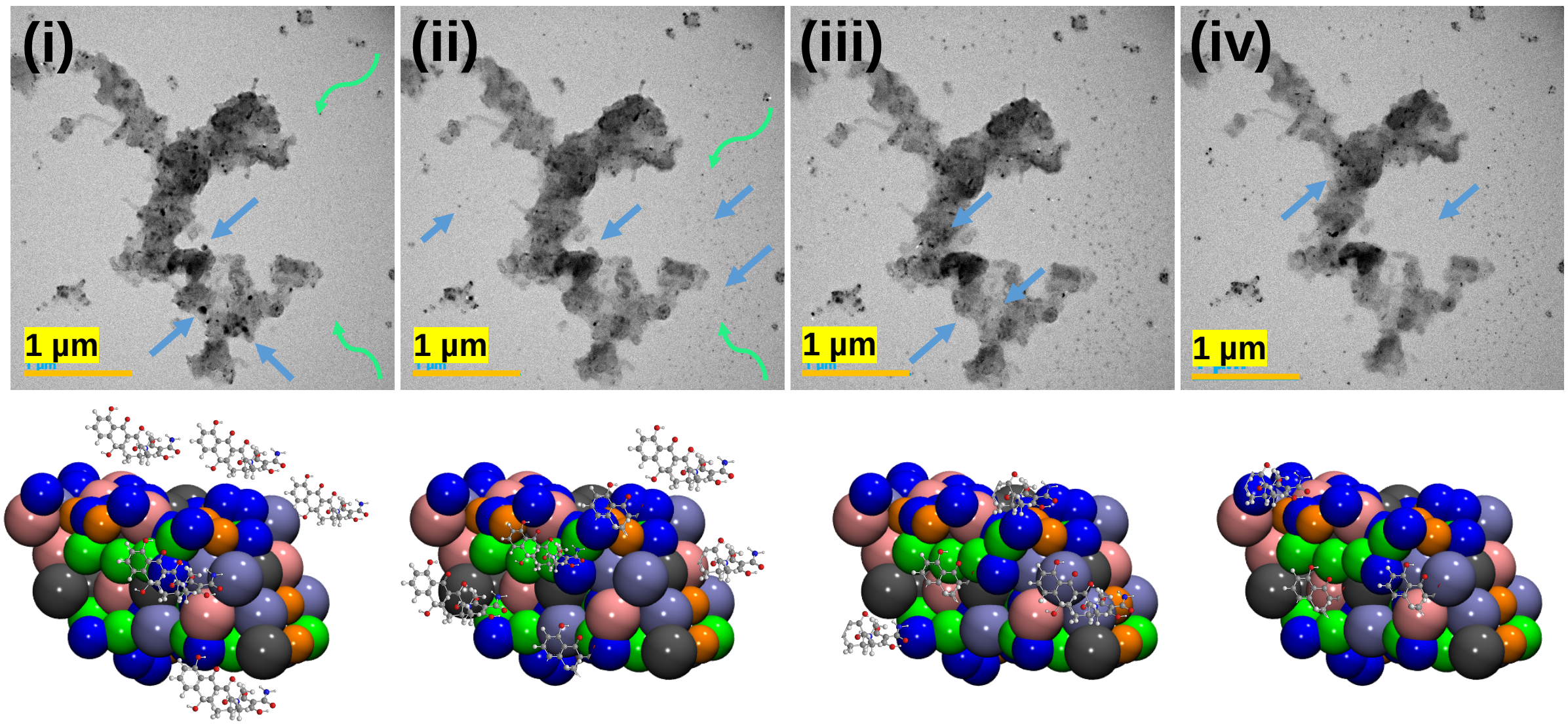

**Fig. 8:** *In situ* LCTEM images and schematic depicting the real-time interaction between (MnFeCoNiCu)N HEN-1:5 NPs and TCL molecules. The flow of liquid TCL is indicated by bright green arrows, and the blue arrows indicate TCL molecules bonded to HEN-1:5 NPs.

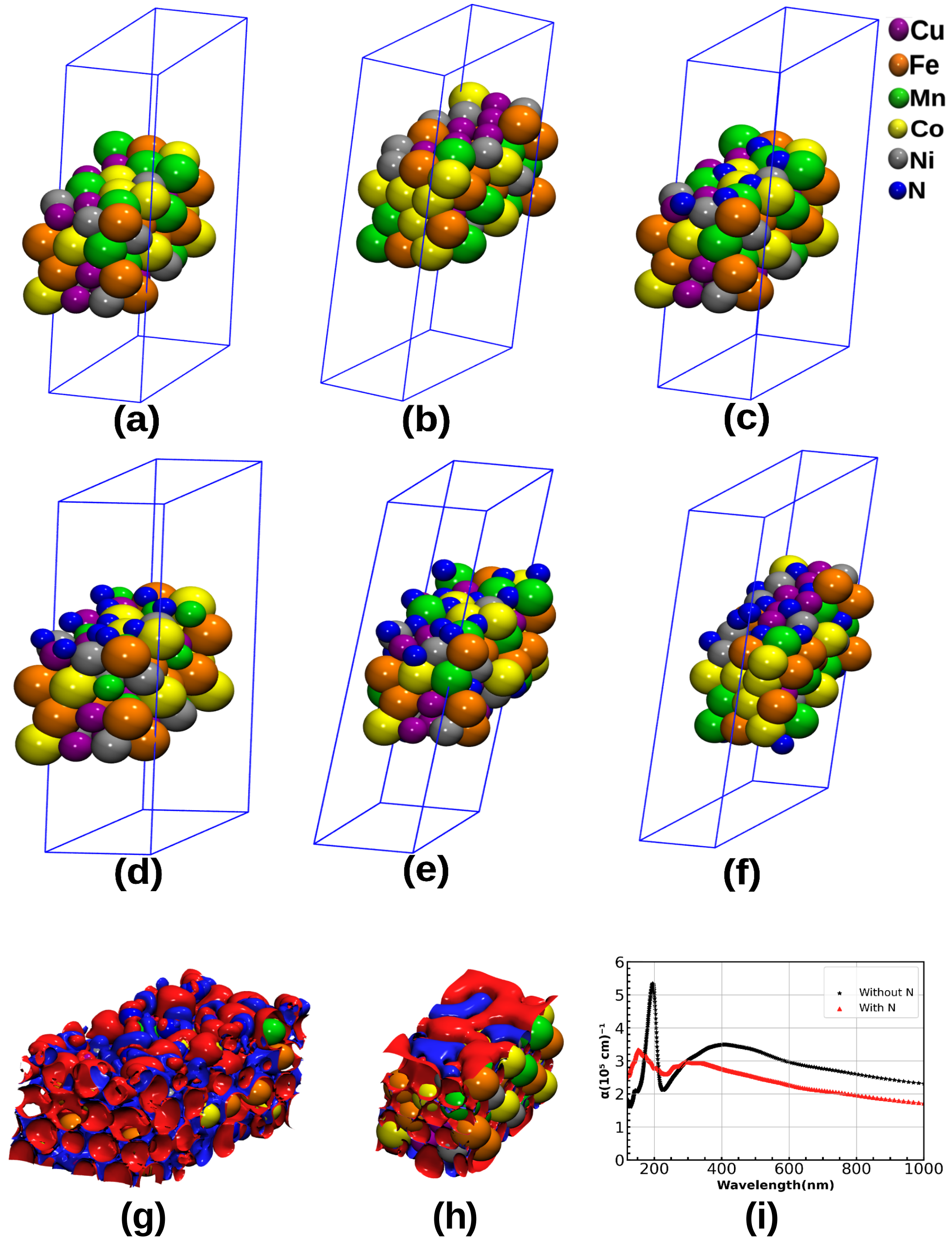

**Fig. 9:** Most energetically stable FeNiMnCoCu HEA configurations (a-b) and the subsequent nitrogenation process. Panels (c) and (d) illustrate two intermediate states of the nitrogen-aggregation high-entropy alloy, while (e) and (f) depict the final HEN structure following the incorporation of 33 nitrogen atoms. Atomic species are represented as Cu (purple), Mn (green), Co (yellow), Ni (silver), Fe (orange), and N (blue). Spin density variation and optical response of the HEN system. Panel (g) displays the spin density distribution on the HEA surface prior to nitrogenation, with blue and red regions representing negative and positive spin density accumulation, respectively. Panel (h) depicts the corresponding spin density variation for the final HEN configuration. Panel (i) provides a comparison of the absorption coefficients for both the pristine HEA and the HEN structures.

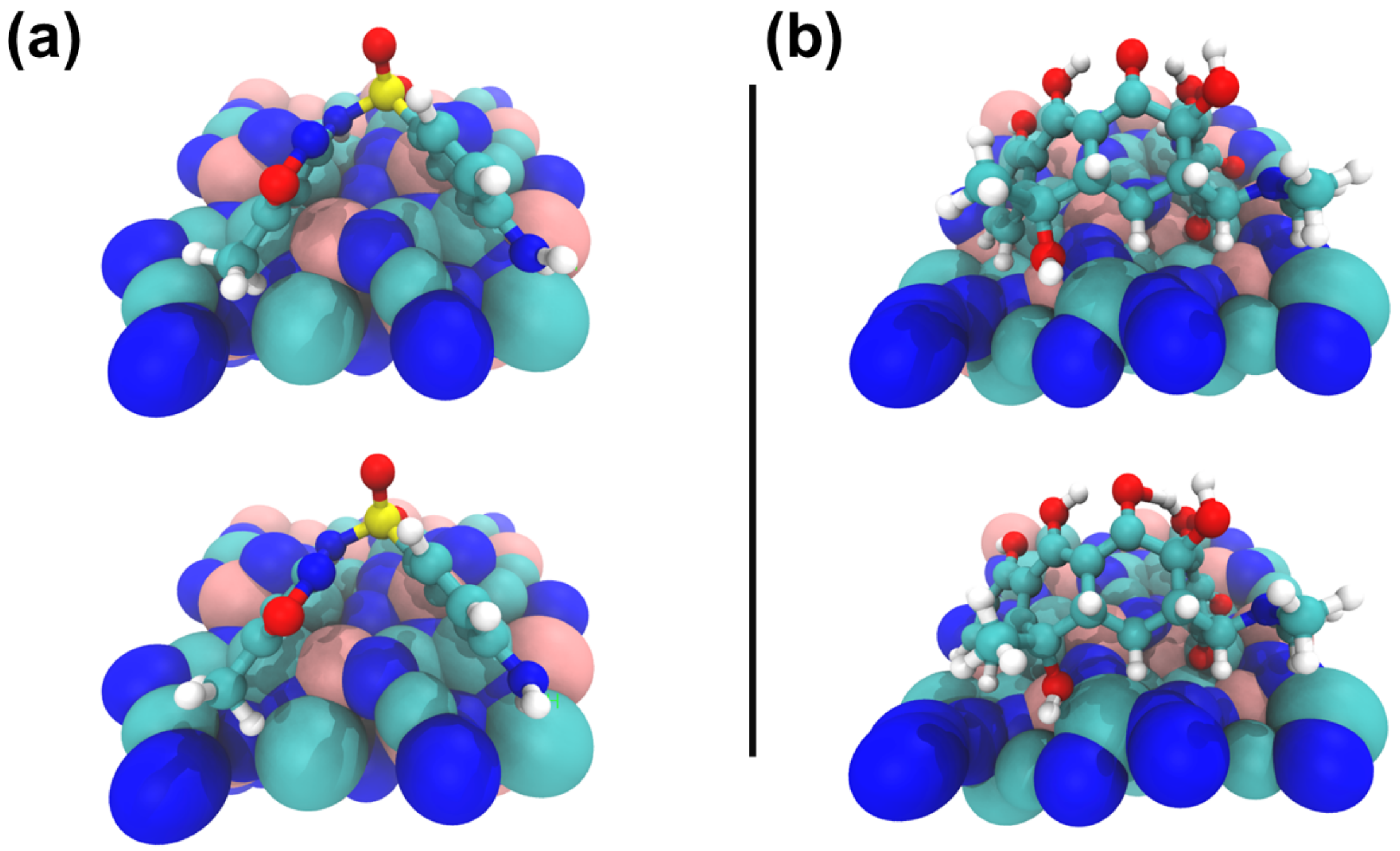

**Fig. 10:** (a) Initial (top) and optimized (bottom) configurations of SME adsorbed on the HEN surface. (b) Initial (top) and optimized (bottom) configurations of TCL adsorbed on the HEN surface.

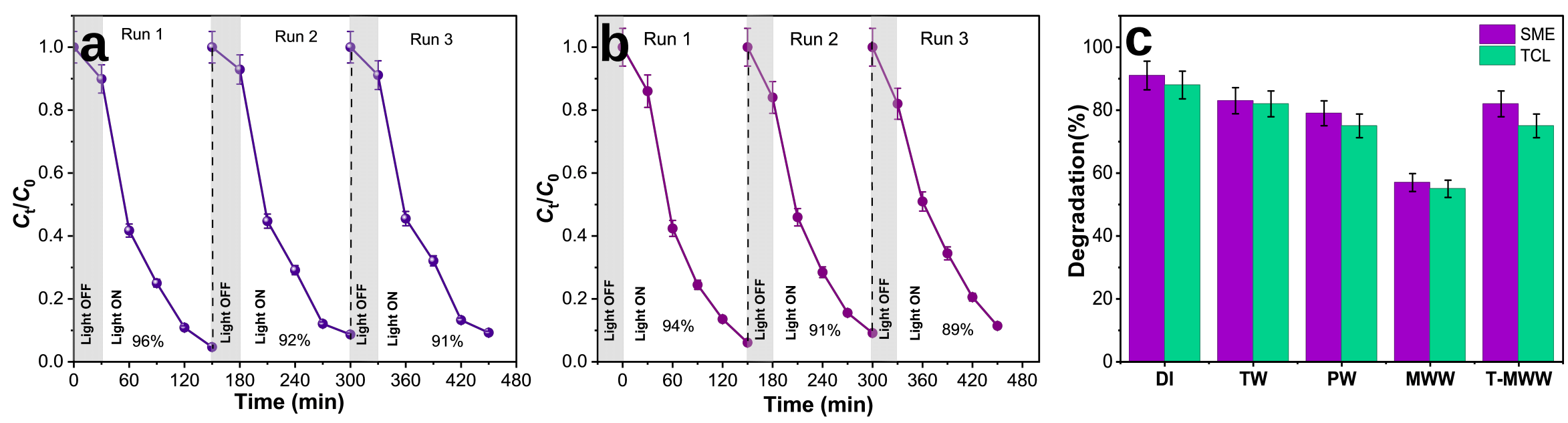

**Fig. 11:** Reusability of (MnFeCoNiCu)N HEN-1:5 NPs for photocatalytic degradation of (a) SME and (b) TCL under visible-light irradiation (experimental conditions: initial antibiotic concentration = 5 mg $L^{-1}$; photocatalyst dose = 0.50 g $L^{-1}$; temperature = 25 °C). (c) Effect of different water matrices on the photocatalytic degradation of SME and TCL over (MnFeCoNiCu)N HEN-1:5 NPs under visible-light irradiation (experimental conditions: initial antibiotic concentration = 5 mg $L^{-1}$; photocatalyst dose = 0.50 g $L^{-1}$; temperature = 25 °C).